\documentclass[bib]{ba}

\usepackage{graphicx}
\usepackage{bayes}
\usepackage{epstopdf}
\usepackage{amssymb}
\usepackage{amsmath}
\usepackage{color}
\usepackage{moreverb}
\usepackage{amssymb}
\usepackage{makeidx}
\usepackage{amsfonts}
\usepackage{subfigure}
\usepackage{algorithm}
\usepackage{footmisc}
\usepackage{units}
\usepackage{hyperref}

\usepackage{tikz}
\usetikzlibrary{arrows,calc}
\tikzset{
link/.style={->},
line/.style={thick},
line2/.style={dashed},
trans/.style={thick,->,shorten >=0.2},
}

\providecommand{\A}{\mathcal{A}}

\providecommand{\C}{\mathcal{C}}

\providecommand{\N}{\mathcal{N}}

\renewcommand{\S}{\mathbb{S}}
\providecommand{\cM}{{\cal{M}}}

\DeclareMathOperator{\Beta}{Beta}

\graphicspath{{./}{./figures/}{../figures/}}

\def\Polya{P\'olya }

\begin{document}

\inserttype[ba0001]{article}
\renewcommand{\thefootnote}{\fnsymbol{footnote}}
\author{S. Filippi and C. Holmes}{
  \fnms{Sarah}
  \snm{Filippi}
  \footnotemark[1]\ead{filippi@stats.ox.ac.uk}
  and
 \fnms{Chris C.}
 \snm{Holmes}
 \footnotemark[2]\ead{cholmes@stats.ox.ac.uk}
}

\title[small title]{A Bayesian nonparametric approach to  testing for dependence between random variables}

\maketitle

\footnotetext[1]{
 Department of Statistics, University of Oxford, England, 
 \href{mailto:filippi@stats.ox.ac.uk}{filippi@stats.ox.ac.uk}
}
\footnotetext[2]{
 Department of Statistics and Oxford-Man Institute, University of Oxford, England, 
 \href{mailto:cholmes@stats.ox.ac.uk}{cholmes@stats.ox.ac.uk}
}

\renewcommand{\thefootnote}{\arabic{footnote}}

\begin{abstract}

Nonparametric and nonlinear measures of statistical dependence between pairs of random variables are important tools in modern data analysis. In particular the emergence of large data sets can now support the relaxation of linearity assumptions implicit in traditional association scores such as correlation. Here we describe a Bayesian nonparametric procedure that leads to a tractable, explicit and analytic quantification of the relative evidence for dependence vs independence. Our approach uses \Polya tree priors on the space of probability measures which can then be embedded within a decision theoretic test for dependence.  \Polya tree priors can accommodate known uncertainty in the form of the underlying sampling distribution and provides an explicit posterior probability measure of both dependence and independence. Well known advantages of having an explicit probability measure include: easy comparison of evidence across different studies; encoding prior information; quantifying changes in dependence across different experimental conditions, and; the integration of results within formal decision analysis.

\end{abstract}

\section{Introduction}

Quantifying the evidence for dependence or testing for departures from independence between  random variables is an increasingly important  task and has been the focus of a number of studies in the past decade.  A typical motivating example comes from the field of biology where a growing abundance of genetic, proteomic and transcriptomic data is being produced. In order to unravel the existing relationships between different molecular species (genes, proteins, ...) involved in a biological system, large datasets are commonly screened for evidence of association between the pairs of variables. This requires adequate statistical procedures to quantify the evidence of dependence (or lack of independence) between two samples of typically continuous random variables.
\par
In this article, we propose a Bayesian nonparametric procedure to derive a probabilistic measure of dependency between two samples $x$ and $y$ without assuming a known form for the underlying distributions. In particular let ${\cal{M}}_0$ denote a model, or hypothesis, of independence and ${\cal{M}}_1$ a model, or hypothesis, of dependence. The posterior probability, $p({\cal{M}}_1|x,y)$, is then a natural measure of 
the strength of evidence for dependence between the two samples against independence.
The Bayes Factor quantifying the relative evidence in the data in favour of ${\cal{M}}_1$ over ${\cal{M}}_0$ is simply,
$$
BF = \frac{p(x, y | {\cal{M}}_1)}{p(x | {\cal{M}}_0) p(y | {\cal{M}}_0)} ,
$$
which is the ratio of the prior predictive probability of the observed data given the two competing hypotheses.
This Bayes Factor implicitly avoids conditioning on the form of the unknown distribution functions. The role of Bayesian nonparametrics is to allow one to accommodate this uncertainty via a prior measure on the space of probability measures, for instance,
$$
p(x, y | {\cal{M}}_1) = \int f(x, y | {\cal{M}}_1) \pi(d F | {\cal{M}}_1)
$$
where $\pi(\cdot)$ is a Bayesian nonparametric prior with wide support over the space of probability measures on the joint sample space $\Omega_X \times \Omega_Y$; see for example \cite{hjort2010bayesian}.

We use \Polya tree priors \citep{lavine1992some,mauldin1992polya,lavine1994more} to model the unknown distributions of the data. We show that the use of such priors leads to an analytic derivation of our association measure $p({\cal{M}}_1|x,y)$. In particular, this measure of dependence involves a finite analytic calculation though the \Polya tree prior is defined over an infinite recursive partition. 
\Polya tree priors have previously been used to derive Bayesian nonparametric procedure for two sample hypothesis testing~\citep{holmes2009two,ma2011coupling}  and extensions of these priors have been proposed to model distributions indexed by covariates~\citep{trippa2011multivariate}.  The ``two-sample testing'' problem is different to that considered here in that it considers the same measurement, or outcome, $Y$, measured on independent samples under different 
conditions and tests for evidence of the  ``treatment'' or covariate effect, whereas our paper is concerned with exploring evidence for statistical association between two joint measurement variables, $\{Y, X\}$, recorded together on a set of samples. However, our approach exploits a similar framework to the testing procedure from \cite{holmes2009two}. In particular, our association measure necessitates the construction of \Polya tree priors on a two-dimensional space, and as discussed at the end of the paper, this engenders new challenges regarding the partitioning scheme to be employed. It is worth noting that \Polya trees offer a more appropriate nonparametric model than say Bayesian histograms, with Dirichlet priors \citep{leonard1973bayesian}, as the recursive tree structure of the \Polya tree is indexed on the measurement variable, whereas the Dirichlet prior for histograms is for unordered categorical data and local dependence between measurement bins must be introduced via a hierarchical prior.  Moreover the \Polya tree is defined via an infinite sequence partitioning, bypassing the need to truncate at some level, and, as noted above, our approach can compute the Bayes factors from the infinite sequence.

\par
Numerous frequentist approaches have been developed for identifying associations between two samples \citep{shannon1949mathematical,cover1991elements,reshef2011detecting,gretton2010consistent} but to the best of our knowledge this is the first Bayesian nonparametric procedure  to quantify of the relative evidence of dependence vs independence. Being able to provide an explicit probability of dependence is attractive for a number of reasons. First, it can be combined with a variety of probabilistic approaches. In particular, it can be easily merged with the decision theory framework in order for optimal statistical decisions to be made in the face of uncertainty. Another important property of probabilistic measures is their interpretability. Indeed, a given level $p=p({\cal{M}}_1|x,y)$ of this measure is exactly the probability of a dependent generative model given the data and the probability of an independent generative model is simply $1-p$. Over and above the standard arguments in favour of Bayesian inference, one explicit consequence of the coherence is that we can explicitly quantify the evidence for a change in dependence between two variables across two or more conditions. For example, if there is evidence that two dependent variables $\{X, Y\}$ become independent on application of a treatment, or across disease states. Answering such questions is problematic from a non-Bayesian perspective, as a null hypothesis of dependence is of higher dimension than the corresponding alternative hypothesis of independence which is nested {\em{under}} the null. This makes the quantification of a p-value extremely challenging. In  Bayesian analysis, the symmetry of the model space makes for a simple and intuitive solution. In Section 4 we illustrate this issue using an important application in cancer genetics.

\par
The remainder of the paper is organised as follows. We first introduce the \Polya tree priors in Section~\ref{sec:Polya} and summarise their main properties. In section 3, we describe our nonparametric procedure to test for dependencies between two samples. We then illustrate in section 4 our approach on data generated from simple models and then we apply our procedure to two real world problems from biology. In the Appendix we provide an empirical calibration comparing our method to that of other non-Bayesian approaches in the literature. 

\section{Polya Tree priors}\label{sec:Polya}
\Polya trees form a class of distributions for random probability measures $F$ on some domain $\Omega$ \citep{ferguson1974prior} by considering a recursive partition of $\Omega$ into disjoint measurable spaces and constructing random measures on each of these spaces. A binary recursive partitioning is typically used for one-dimensional domains: $\Omega$ is divided in two disjoint sets $C_0$ and $C_1$ which themselves are divided in two other disjoints sets $C_0=C_{00}\cup C_{01}$ and $C_1=C_{10}\cup C_{11}$, and so on. The infinite recursive partition is denoted by $\C=\{C_j, j=0,1,00,01,10,11,\dots \}$; the partition at level $k$ is comprised of $2^k$ sets $C_j$ where $j$ are all binary sequences of length $k$. 
\par
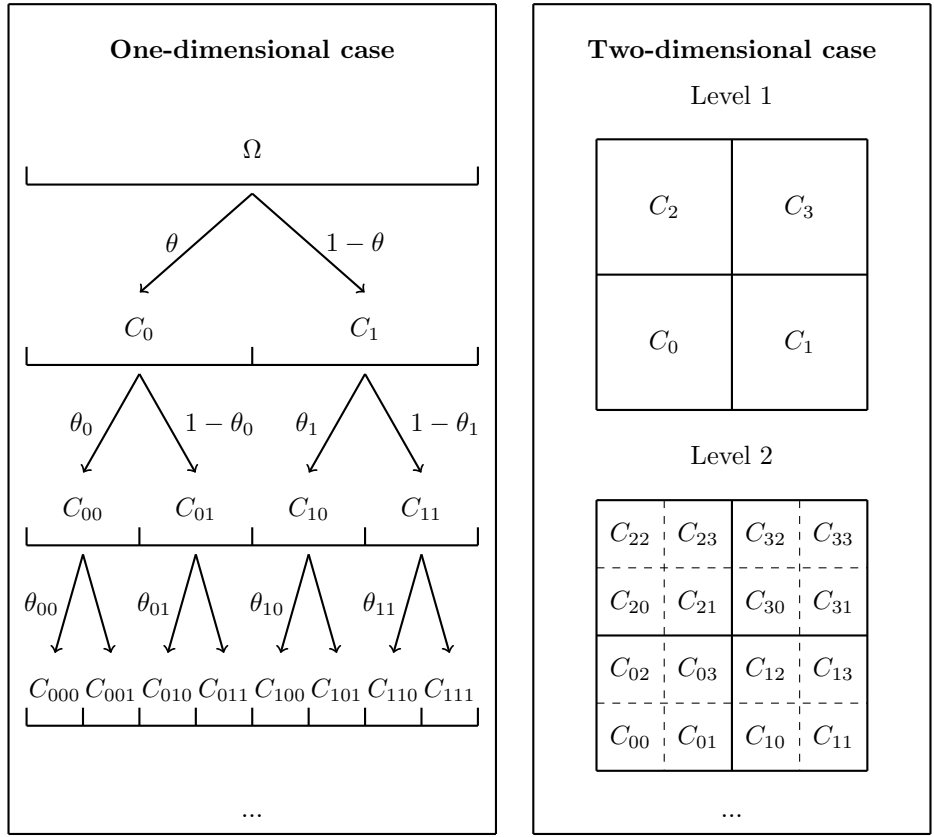
\begin{figure}
 \begin{tikzpicture}[scale=1.2]
 
  \draw[line] (-0.2,0.8) -- (5.2,0.8); 
  \draw[line] (-0.2,0.8) -- (-0.2,10); 
   \draw[line] (-0.2,10) -- (5.2,10); 
    \draw[line] (5.2,10) -- (5.2,0.8); 
    
    \node at (2.5, 9.5) {\bf One-dimensional case};
 
 \draw[line] (0,8) -- (0,8.2); 
 \draw[line] (5,8) -- (5,8.2); 
 \draw[line] (0,8) -- (5,8);
 \node at (2.5,8.4) {$\Omega$};
 
 \draw[line] (0,6) -- (5,6);
 \draw[line] (0,6) -- (0,6.2); 
 \draw[trans] (2.5,7.9) -- (1.25,6.8) node[midway,left] {$\;\theta\;$};
  \node at (1.25,6.4) {$C_0$};
 \draw[line] (2.5,6) -- (2.5,6.2); 
  \draw[trans] (2.5,7.9) -- (3.75,6.8) node[midway,right] {$\;1-\theta\;$};
 \node at (3.75,6.4) {$C_1$};
 \draw[line] (5,6) -- (5,6.2); 
 
 \draw[line] (0,4) -- (5,4);
 \draw[line] (0,4) -- (0,4.2); 
 \draw[trans] (1.25,5.9) -- (0.625,4.8) node[midway,left] {$\;\theta_{0}\;$};
   \node at (0.625,4.4) {$C_{00}$};
  \draw[line] (1.25,4) -- (1.25,4.2); 
  \draw[trans] (1.25,5.9) -- (1.875,4.8) node[midway,right] {$\;1-\theta_{0}\;$};
    \node at (1.875,4.4) {$C_{01}$};
 \draw[line] (2.5,4) -- (2.5,4.2); 
  \draw[trans] (3.75,5.9) -- (3.125,4.8) node[midway,left] {$\;\theta_{1}\;$};
  \node at (3.125,4.4) {$C_{10}$};
  \draw[line] (3.75,4) -- (3.75,4.2);
    \draw[trans] (3.75,5.9) -- (4.3755,4.8) node[midway,right] {$\;1-\theta_{1}\;$};
    \node at (4.375,4.4) {$C_{11}$}; 
 \draw[line] (5,4) -- (5,4.2);  
 
  \draw[line] (0,2) -- (5,2);
 \draw[line] (0,2) -- (0,2.2); 
  \draw[trans] (0.625,3.9) -- (0.625/2,2.8) node[midway,left] {$\;\theta_{00}$};
 \node at (0.625/2,2.4) {$C_{000}$};
  \draw[line] (0.625,2) -- (0.625,2.2); 
    \draw[trans] (0.625,3.9) -- (0.625+0.625/2,2.8);
  \node at (0.625+0.625/2,2.4) {$C_{001}$};
  \draw[line] (1.25,2) -- (1.25,2.2); 
  \draw[trans] (1.875,3.9) -- (1.25+0.625/2,2.8) node[midway,left] {$\;\theta_{01}$};
  \node at (1.25+0.625/2,2.4) {$C_{010}$};
   \draw[line] (1.875,2) -- (1.875,2.2);
    \draw[trans] (1.875,3.9) -- (1.875+0.625/2,2.8);
   \node at (1.875+0.625/2,2.4) {$C_{011}$}; 
 \draw[line] (2.5,2) -- (2.5,2.2); 
  \draw[trans] (3.125,3.9) -- (3.125-0.625/2,2.8) node[midway,left] {$\;\theta_{10}$};
 \node at (2.5+0.625/2,2.4) {$C_{100}$};
  \draw[line] (3.125,2) -- (3.125,2.2);
  \draw[trans] (3.125,3.9) -- (3.125+0.625/2,2.8);
  \node at (3.125+0.625/2,2.4) {$C_{101}$};  
  \draw[line] (3.75,2) -- (3.75,2.2); 
   \draw[trans] (4.375,3.9) -- (4.375-0.625/2,2.8) node[midway,left] {$\;\theta_{11}$};
  \node at (3.75+0.625/2,2.4) {$C_{110}$};
  \draw[line] (4.375,2) -- (4.375,2.2); 
  \draw[trans] (4.375,3.9) -- (4.375+0.625/2,2.8);
  \node at (4.375+0.625/2,2.4) {$C_{111}$};
 \draw[line] (5,2) -- (5,2.2);  
 
 \node at (2.5,1) {...};
 
\end{tikzpicture}
\quad
 \begin{tikzpicture}[scale=1.2]
 
  \draw[line] (-0.2,0.8) -- (4.2,0.8); 
  \draw[line] (-0.2,0.8) -- (-0.2,10); 
   \draw[line] (-0.2,10) -- (4.2,10); 
    \draw[line] (4.2,10) -- (4.2,0.8); 

    \node at (2, 9.5) {\bf Two-dimensional case};
    
     \node at (2, 9) { Level 1};
     \coordinate (L1) at (0.5,8.5);
     \coordinate (L2) at (3.5,8.5);
	\coordinate (L4) at (0.5,5.5);
	\coordinate (L3) at (3.5,5.5);
	\draw[line] (L1) -- (L2);
	\draw[line] (L2) -- (L3);
	\draw[line] (L3) -- (L4);
	\draw[line] (L4) -- (L1);
	
     \coordinate (L1) at (2,8.5);
     	\coordinate (L2) at (2,5.5);
     \coordinate (L3) at (0.5,7);
	\coordinate (L4) at (3.5,7);
	\draw[line] (L1) -- (L2);
	\draw[line] (L3) -- (L4);
	
\node at (1.25,7.75) {$C_2$};
\node at (2.75,7.75) {$C_3$};	
\node at (1.25,6.25) {$C_0$};
\node at (2.75,6.25) {$C_1$};	
	
     \node at (2, 5) { Level 2};
     \coordinate (L1) at (0.5,4.5);
     \coordinate (L2) at (3.5,4.5);
	\coordinate (L4) at (0.5,1.5);
	\coordinate (L3) at (3.5,1.5);
	\draw[line] (L1) -- (L2);
	\draw[line] (L2) -- (L3);
	\draw[line] (L3) -- (L4);
	\draw[line] (L4) -- (L1);
		
     \coordinate (L1) at (2,4.5);
     	\coordinate (L2) at (2,1.5);
     \coordinate (L3) at (0.5,3);
	\coordinate (L4) at (3.5,3);
	\draw[line] (L1) -- (L2);
	\draw[line] (L3) -- (L4);		
	
	    \coordinate (H) at (1.25,4.5);
	     \coordinate (B) at (1.25,1.5);
	\draw[line2] (H) -- (B);
		    \coordinate (H) at (2.75,4.5);
	     \coordinate (B) at (2.75,1.5);
	\draw[line2] (H) -- (B);
	    \coordinate (L) at (0.5,2.25);
	     \coordinate (R) at (3.5,2.25);
	\draw[line2] (L) -- (R);	
		    \coordinate (L) at (0.5,3.75);
	     \coordinate (R) at (3.5,3.75);
	\draw[line2] (L) -- (R);	
	
\node at (0.875,4.125) {$C_{22}$};
\node at (0.875+0.75,4.125) {$C_{23}$};
\node at (0.875+2*0.75,4.125) {$C_{32}$};
\node at (0.875+3*0.75,4.125) {$C_{33}$};

\node at (0.875,4.125-0.75) {$C_{20}$};
\node at (0.875+0.75,4.125-0.75) {$C_{21}$};
\node at (0.875+2*0.75,4.125-0.75) {$C_{30}$};
\node at (0.875+3*0.75,4.125-0.75) {$C_{31}$};	

\node at (0.875,4.125-2*0.75) {$C_{02}$};
\node at (0.875+0.75,4.125-2*0.75) {$C_{03}$};
\node at (0.875+2*0.75,4.125-2*0.75) {$C_{12}$};
\node at (0.875+3*0.75,4.125-2*0.75) {$C_{13}$};	

\node at (0.875,4.125-3*0.75) {$C_{00}$};
\node at (0.875+0.75,4.125-3*0.75) {$C_{01}$};
\node at (0.875+2*0.75,4.125-3*0.75) {$C_{10}$};
\node at (0.875+3*0.75,4.125-3*0.75) {$C_{11}$};	
	
 \node at (2,1) {...};	
\end{tikzpicture}
\caption{{\bf (Left)} Construction of a \Polya tree distribution in the uni-dimensional case: at each junction $j$ the particle has a random probability $\theta_j$ to choose the left branch and $1-\theta_j$ to choose the right one. {\bf (Right)} Illustration of the first two levels of the quadrant partitioning scheme in the two-dimensional case. }
\label{fig:partition}
\end{figure}

To better understand the probability measure constructed on these nested partitions, one can think of a particle going down the tree shown in Figure \ref{fig:partition} Left; at each junction $j$, usually represented in binary format, the particle has a random probability $\theta_j$ to choose the left branch. In \Polya trees, the random branching probability follows a Beta distribution, with $\theta_j\sim \text{Beta}(\alpha_{j,(0)}, \alpha_{j,(1)})$. Given the partition $\C$, the sequence of non-negative vectors $\A=\{\alpha_{j,(0)}, \alpha_{j,(1)}\}_j$ and the sequence of realisations of the random branching variables $\Theta=\{\theta_j\}_j$, it is possible to compute the likelihood for any set of observations $x$:
\begin{equation}
p(x|\Theta,\C,\A)=\prod_j \theta_{j}^{n_{j0}}(1-\theta_{j})^{n_{j1}}
\label{eq:probax_condt}
\end{equation}
where the product is over the set of all partitions and $n_{j0}$ and $n_{j1}$ denote the number of observations that lie in the partitions $C_{j0}$ and $C_{j1}$ respectively. The Beta prior on the partition probability is conjugate to the Binomial likelihood and, integrating out $\theta_j$ for all $j$, we obtain that
\begin{equation}
p(x|\C,\A)=\prod_j \frac{B(n_{j0}+\alpha_{j,(0)},n_{j1}+\alpha_{j,(1)})}{B(\alpha_{j,(0)},\alpha_{j,(1)})}
\label{eq:probax}
\end{equation}
where $B(.,.)$ refers to the Beta function. For more details on \Polya Tree priors, we refer the reader to \cite{ferguson1974prior,lavine1992some,mauldin1992polya,lavine1994more,ghosh2003bayesian,wong2010optional}.
\par
In this paper, we are interested in testing independence between two samples $x$ and $y$. We therefore need to consider the joint space $\Omega_X\times\Omega_Y$ of the two sample spaces. For reasons that will become obvious later on, we recursively subdivide this space into four rectangular regions. We start with partitioning $\Omega_X\times\Omega_Y$ in $4$ quadrants, $\Omega_X\times\Omega_Y=C_0\cup C_1\cup C_2 \cup C_3$, and continue with nested partitions defined by $C_j=C_{j0}\cup C_{j1}\cup C_{j2} \cup C_{j3}$ for any base $4$ number $j$. Thus the partition at level $k$ is formed of $4^k$ sets $C_j$ where $j$ are all quaternary sequences of length $k$. We assume that the sets $C_{j}$ are rectangular, i.e. can be written as a Cartesian product $D\times E$ where $D\subset \Omega_X$ and $E\subset \Omega_Y$. We arbitrarily choose to denote $C_{j0}$ the left bottom region of the set $C_j$, $C_{j1}$ the right bottom region, $C_{j2}$ the left top region, and $C_{j3}$ the right top region for all $j$. This recursive partition $\C=\{C_j, j=0,1,2,3,01,02,03,11,\dots\}$ is illustrated in Figure \ref{fig:partition}B. Similarly to the one-dimensional case, a probability measure can be constructed on this recursive partition by defining random branching probabilities in the recursive quaternary partition. In the following we will use different distributions for these branching probabilities.

\section{A Bayesian nonparametric measure of dependence}

\subsection{The approach}

Given a $N$ sample $(x,y)$ which are i.i.d. realisations of the random vector $(X,Y)$, we wish to evaluate the evidence for the competing hypotheses:
\begin{align*}
{\cal{M}}_0:& \text{ X and Y are independent random variables};\\
{\cal{M}}_1:& \text{ X and Y are dependent random variables}.\;.
\end{align*}
We denote by $F_{XY}$ the unknown joint probability distribution of $(X,Y)$ and by $F_X$ and $F_Y$ the two unknown marginal distributions. Following a Bayesian approach, we aim at estimating the posterior probability 
\[ p({\cal{M}}_1|x,y)\propto p(x,y|\cM_1) p(\cM_1)\]
where $p(\cM_1)$ represents prior beliefs regarding the competing hypotheses. We specify our uncertainty in the distribution $F_{XY}$ via a \Polya tree prior. Denoting by $\Omega_X$ and $\Omega_Y$ the domains of the probability measures $F_X$ and $F_Y$ respectively, we consider a recursive quaternary partition of $\Omega_X\times\Omega_Y$ into disjoint measurable sets as described in the previous section. 

\par
Under model $\cM_0$, we assume that samples $x$ and $y$ are independent. We can therefore think of the partitioning in terms of x-axis and y-axis separately. Let $\xi_{j,X}$ and $\xi_{j,Y}$ denote the independent random branching probabilities which determine the probability of going in the ``left" region of $C_j$ (i.e. $C_{j0}\cup C_{j2}$) and the``bottom" region of $C_j$ (i.e. $C_{j0}\cup C_{j1}$) respectively. Similarly to the one-dimensional case, we assume that the random branching probabilities follow Beta distributions,  $\xi_{j,X}\sim\Beta(\alpha_{j,X,(0)},\alpha_{j,X,(1)})$ and $\xi_{j,Y}\sim\Beta(\alpha_{j,Y,(0)},\alpha_{j,Y,(1)})$.
By independence of $\xi_{j,X}$ and $\xi_{j,X}$, the likelihood of the data given the partition $\C$,  the sequence of random branching variables $\Xi=\{\xi_{j,X},\xi_{j,Y}\}_j$,  $\A_X=\{\alpha_{j,X,(0)},\alpha_{j,X,(1)}\}_j$ and $\A_Y=\{\alpha_{j,Y,(0)},\alpha_{j,Y,(1)}\}_j$ can be computed as follows
$$p(x,y|\Xi,\C,\A_X,\A_Y, \cM_0)=\prod_j \xi_{j,X}^{n_{j0}+n_{j2}} (1- \xi_{j,X})^{n_{j1}+n_{j3}}  \xi_{j,Y}^{n_{j0}+n_{j1}} (1- \xi_{j,Y})^{n_{j2}+n_{j3}}$$
where, for quaternary sequence $j\in\{0,1,2,3,01,\dots,03,\dots,31,\dots, 33,\dots\}$, $n_j$ is the number of observations falling in $C_j$.  Integrating the random branching probabilities out, we have
\begin{align}
p(x,y|\C,\A_X,\A_Y,\cM_0)&=\prod_j  \frac{B(n_{j0}+n_{j2}+\alpha_{j,X,(0)},n_{j1}+n_{j3}+\alpha_{j,X,(1)})}{B(\alpha_{j,X,(0)},\alpha_{j,X,(1)})}\nonumber\\
&\qquad\qquad\times\frac{ B( n_{j0}+n_{j1}+\alpha_{j,Y,(0)}, n_{j2}+n_{j3}+\alpha_{j,Y,(1)}) }{B(\alpha_{j,Y,(0)},\alpha_{j,Y,(1)})}\;.
\label{eq:pxyH0}
\end{align}

\par
Under the $\cM_1$ hypothesis, we do not assume independence between samples $x$ and $y$. In this case, for each set $j$, the random branching probability $\theta_j=(\theta_{j,(0)},\theta_{j,(1)},\theta_{j,(2)},\theta_{j,(3)})$ is a random vector taking values in the simplex $\S^3$. For every $i\in\{0,1,2,3\}$, $\theta_{ji}$ is the probability that the particle falls in the quadrant $C_{ji}$.
We use Dirichlet distributions with parameters $\alpha_j=(\alpha_{j,(0)},\alpha_{j,(1)},\alpha_{j,(2)},\alpha_{j,(3)})$ for the random branching variables $\{\theta_j\}_j$ \citep{hanson2006inference}. Hence the marginal likelihood is
\begin{equation}
p(x,y|\C,\A,\cM_1)=\prod_j\frac{\tilde B(\tilde{n}_j+\alpha_j)}{\tilde B(\alpha_{j})}
\label{eq:pxyH1}
\end{equation}
where  $\tilde{n}_j=(n_{j0},n_{j1},n_{j2},n_{j3})$ and $\tilde B$ is the multinomial Beta function defined as
$$\tilde B(\alpha_j)=\frac{\prod_{i=0}^3\Gamma(\alpha_{j,(i)})}{\Gamma(\sum_{i=0}^3\alpha_{j,(i)})}$$ where $\Gamma$ designates the Gamma function.
\par
Typically the values for the $\alpha_{j,(i)}$ are of the form $ck^2$ for $\alpha$ parameters at level $k$ \citep{walker1999bayesian}; we recall that $k$ is the depth of the set $C_j$ and  the length of the quaternary sequence $j$. We will follow this convention so that $a_j=\alpha_{j,(i)}$ for $i\in \{0,1,2,3\}$. In addition, to ensure that the prior distributions are equivalent under the two models, we will assume that, for all $j$, $\alpha_{j,X,(0)}=\alpha_{j,(0)}+\alpha_{j,(2)}=2a_j$, $\alpha_{j,X,(1)}=\alpha_{j,(1)}+\alpha_{j,(3)}=2a_j$, $\alpha_{j,Y,(0)}=\alpha_{j,(0)}+\alpha_{j,(1)}=2a_j$ and $\alpha_{j,Y,(1)}=\alpha_{j,(2)}+\alpha_{j,(3)}=2a_j$. 
\par
To compare evidence in favours of both hypotheses, we compute the following ratio
$$
\frac{p(\cM_0|x,y)}{p(\cM_1|x,y)}=\frac{p(x,y|\cM_0)}{p(x,y|\cM_1)}\frac{p(\cM_0)}{p(\cM_1)} 
$$
where the first term is the Bayes factor which can be written as a product over all partitions:
\begin{equation}
\frac{p(x,y|\cM_0)}{p(x,y|\cM_1)}=\prod_j b_j, \;.
\label{eq:BF}
\end{equation}
where $b_j$ is defined below. From equations~\eqref{eq:pxyH0} and~\eqref{eq:pxyH1} and expressing Beta and multinomial Beta functions in terms of Gamma functions, we have
\begin{align}
b_j&=\frac{\Gamma(n_{j0}+n_{j2}+2a_j)\Gamma(n_{j1}+n_{j3}+2a_j)\Gamma(n_{j0}+n_{j1}+2a_j)\Gamma(n_{j2}+n_{j3}+2a_j)}{\Gamma(n_{j0}+n_{j1}+n_{j2}+n_{j3}+4a_j)\Gamma(n_{j0}+a_j)\Gamma(n_{j1}+a_j)\Gamma(n_{j2}+a_j)\Gamma(n_{j3}+a_j)}\nonumber\\
&\qquad \times \frac{\Gamma(4a_j)\Gamma(a_j)^4}{\Gamma(2a_j)^4}\;.\label{eq:bj}
\end{align}
The product in~\eqref{eq:BF} is defined over the infinite set of partitions. However for any set $C_j$ containing zero or one data point (i.e. such that $n_j=n_{j0}+n_{j1}+n_{j2}+n_{j3}\leq 1$ ), $b_j=1$. Therefore, only subsets with at least two data points contribute to this product. The Bayesian measure of  the strength of evidence for dependence between the two samples against independence involves a finite analytic calculation even though~\eqref{eq:BF} is over the infinite number of levels in the tree. 
\par
The procedure is described in Algorithm \ref{algo}. For each set $C_j$ containing more than one datapoint, the term $b_j$ measures the relative evidence in favour of ${\cal{M}}_0$ given the number of datapoints falling in each of the four quadrants of $C_j$. Intuitively, for each set $C_j$ we perform a Bayesian independence test based on the local two-by-two contingency table. \Polya tree priors provides us with a theoretical framework to perform these ``local" independence tests at every level while taking into account potential dependences on neighboring sets. In addition, it allows us to compute the Bayes Factor analytically without having to chose any arbitrary level or any truncation. The parameter $a_j$ which decreases with the depth of the set $C_j$ enables us to give more importance to ``local" independence tests at the higher levels than at the deepest levels. In the next subsection, we investigate the impact of the choice of this parameter.

\begin{algorithm}
\caption{Bayesian nonparametric evidence for independence}
\label{algo}
\begin{enumerate}
\item Fix the quadrant partitioning scheme; choose a constant $c$.

For every set $C_j$ containing more than one data point, compute $b_j$ defined in \eqref{eq:bj} with $a_j=c k^{2}$ where $k$ is the depth of the set $C_j$.

\item Assuming equal prior belief for both hypotheses, $$p(\cM_1|x,y)= \frac{1}{1+\prod_j b_j} = 1 - p(\cM_0|x,y)$$
\end{enumerate}
\end{algorithm}

\subsection{Sensitivity to choice of $\mathcal{A}$}
The proposed procedure relies on a choice of the sequence of non-negative vector $\mathcal{A}$, $\mathcal{A}_X$ and $\mathcal{A}_Y$. As discussed above, the $\alpha$ parameters are constant per level and such that $\alpha_{j,(i)}=ck^2$ where $k$ is the depth of the set $C_j$. In addition, $\alpha_{j,X,(i)}=\alpha_{j,Y,(i)}=2ck^2$. The parameter $c$ controls the speed of divergence of the $\alpha$ parameters with the depth $k$  and therefore the relative contribution of each level of the partition to the Bayes Factor. We have investigated the impact of the setting of $c$ (see Figure S1 and S2 in Supplementary Material) and observe that small values of $c$ typically favours the simpler model $(\cM_0)$ especially when the number of samples is small and there is not enough evidence to determine $\cM_1$. This is to be expected as Bayesian modelling encompasses a natural Occam factor in the prior predictive (see for example chapter 28 of \cite{mackay2003information}). We have found $c=5$ to be a reasonable canonical choice but practitioners are strongly advised to explore the setting for their own analysis.


\subsection{Choice of the partition}\label{sec:partition}

\subsubsection{Basic approach: partition centred on the median of the data }
The inference resulting from a \Polya tree model is known to strongly depend on the specification of the partition $\C$ over the data space \citep{Paddock2003}, and
a multitude of quadrant partitioning scheme could be used in our procedure. As a default, partially subjective approach we suggest to construct a partition based on the quantiles of two normal distributions (for the x- and y-axis respectively). In other words, both variables $x$ and $y$ are transformed through the inverse cumulative distributive function of a normal distribution;  a quaternary recursive partition of $[0,1]\times[0,1]$ is then constructed by subdividing it into four rectangular quadrants of equal size (see Figure~\ref{fig:partitionChoice}, Left) The mean and standard deviation of the normal distributions can be derived from empirical estimates of the location and spread of the two samples. In the next section we use the median and the median absolute deviation as this choice induces robustness to potential outliers. 

\subsubsection{A simple partial optimizing procedure for partition centering}
\begin{figure}
\includegraphics[width=\textwidth]{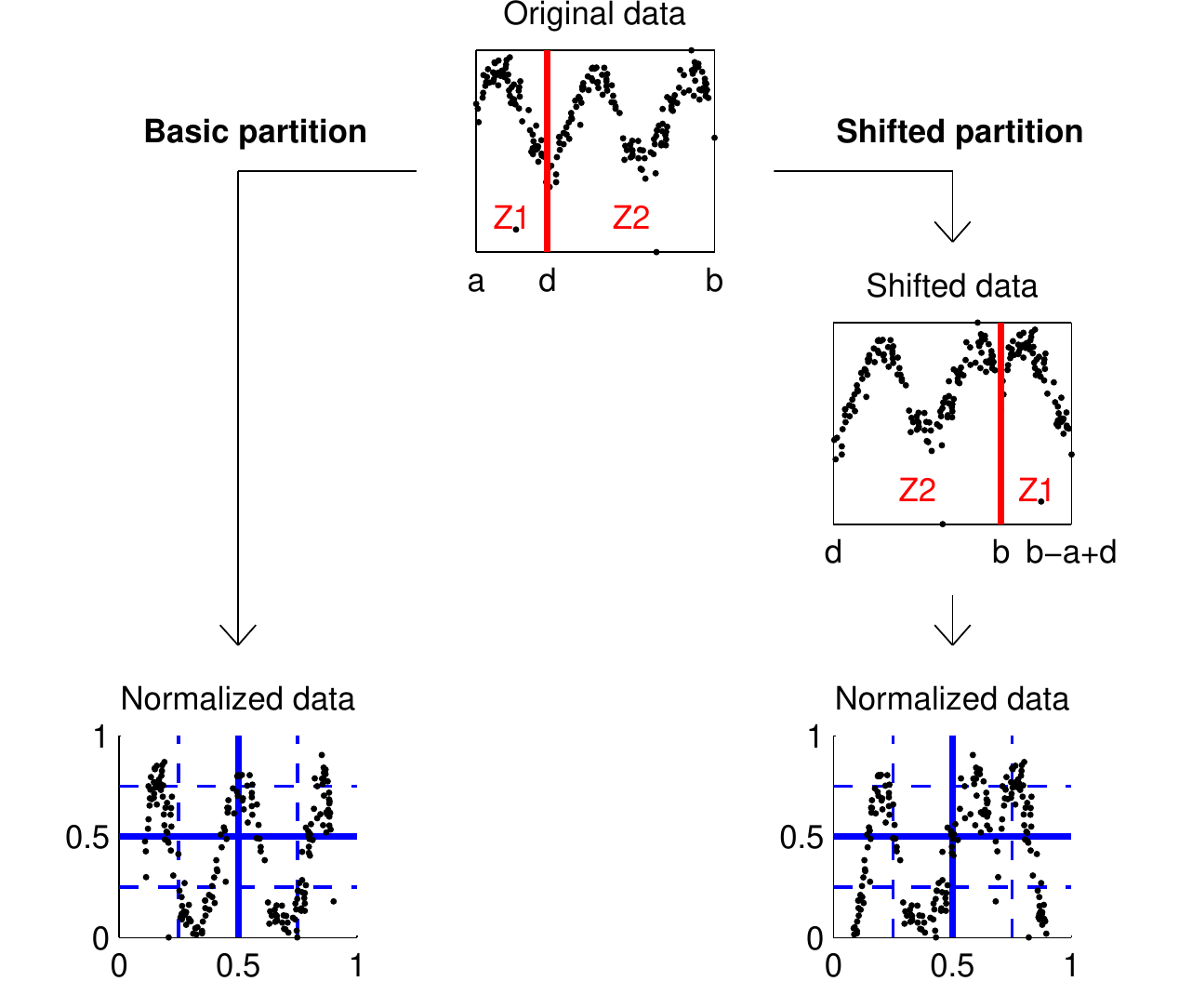}
\caption{{\bf Construction of the partition.} Both the basic approach and the shifted approach are illustrated on a simulated sinusoidal dataset with some outliers. Under the basic approach (Left column), the data are marginally transformed via the inverse of the c.d.f. of normal distributions. The shifted approach consists in shifting the central location of the partition by a factor $\delta$ and wrapping the data around: the data space is divided into two regions $\mathcal{Z}_1=\{(x,y), x\leq \delta\}$ and $\mathcal{Z}_2=\{(x,y), x > \delta\}$ which are then juxtaposed. The obtained ``shifted" data are then normalised  via the inverse of the CDF of a normal distribution. Quaternary recursive partitions of $[0,1]\times[0,1]$ are constructed by subdividing the normalized data into four rectangular quadrants of equal size (Bottom panels).}
\label{fig:partitionChoice}
\end{figure}

Two random variables $X$ and $Y$ are dependent if and only if the distribution of $Y$ conditional on $X\in \mathcal{D}_X$ is different to  the distribution of $Y$ conditional on $X\in \mathcal{D}_X^C$ for some $\mathcal{D}_X$ a compact set of $\Omega_X$ and $\mathcal{D}_X^C$ is its complement.  In the previous paragraph, we suggest centering the partition on the median $(m_x,m_y)$ of the data. For such choice of partition, at the top-level of the \Polya tree our procedure tests whether  the distribution of $Y$ conditional on $X\in \mathcal{D}_X=(-\infty,m_x]$  is equal to the distribution of $Y$ conditional on $X\in \mathcal{D}_X^C$. The procedure performs the test symmetrically on the x-axis and the y-axis. Instead of focusing on the median, it would be more informative to test whether the distribution of $Y$ conditionally on $X\in \mathcal{D}_X$  is equal to the distribution of $Y$ conditionally on $X\in \mathcal{D}_X^C$ for any compact set $ \mathcal{D}_X\in\Omega_X$.

 In this section, we consider a simple partial optimisation of the partition-centering location by considering different compact sets $\mathcal{D}_X$. The approach involves shifting the central location of the partition as defined by the top-level split, and wrapping the data round to maintain balance in the number of points in each region.  Consider a real number $\delta\in [a,b]$ where $a$ and $b$ are respectively the minimum and maximum values of the data on the x-axis. We denote by $\psi_\delta$ the transformation that divides the data space in two regions $\mathcal{Z}_1=\{(x,y), x\leq \delta\}$ and $\mathcal{Z}_2=\{(x,y), x\ge \delta\}$ and juxtaposes them as  illustrated in Figure~\ref{fig:partitionChoice}, Right. More formally,
$\psi_\delta:[a,b]\times\Omega_Y\to [\delta, b-a+\delta]\times\Omega_Y$ such that
$$\psi_\delta(x,y)=\begin{cases}(b-a+x,y)& \text{if }x\leq \delta,\\ (x,y)& \text{otherwise}.
\end{cases}$$
The obtained ``shifted" data are then transformed through the inverse cumulative distributive function of normal distributions and a quaternary recursive partition of $[0,1]\times[0,1]$ is constructed as in the basic approach. We denote the obtained partition by $\mathcal{C}_\delta$.

 We consider optimising  the marginal evidence of dependence $p(\mathcal{M}_1|x,y)$ by maximizing the Bayes factor as defined in equation \eqref{eq:BF} over all the partitions $\mathcal{C}_\delta$ for $\delta\in[a,b]$. The obtained probability of dependence is therefore
 $$p(\cM_1|x,y)= \frac{1}{1+B_\delta}, $$
 where $B_\delta$ designates the Bayes Factor given the partition $\mathcal{C}_\delta$. This approach is called  the ``empirical Bayes approach" in the rest of the paper. Note that optimising the central location of the partition with respect to $p(\mathcal{M}_1|x,y)$ will naturally tend to inflate the evidence for $\mathcal{M}_1$. However, when testing many pairs of random variables for evidence of dependence we are mainly concerned with the ranking of the pairs for further analysis, rather than explicit quantification of the evidence, and partial optimisation of the partition may well help to produce more stable and acurate rankings.

\section{Applications}
In this section, we illustrate the performance of our Bayesian nonparametric procedure for detecting dependence across different datasets. We first test the procedure on simple models proposed by \cite{kinney2014equitability} and then apply it on two real examples from biology.

\subsection{Illustration for simple datasets}
\begin{figure}
\includegraphics{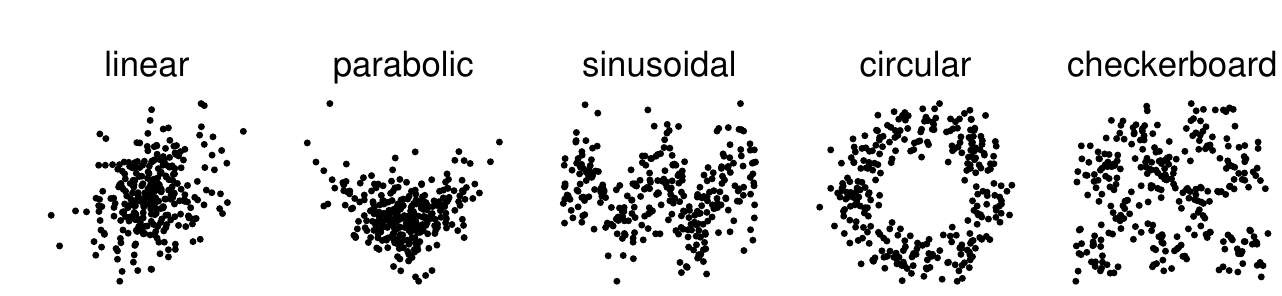}
\includegraphics{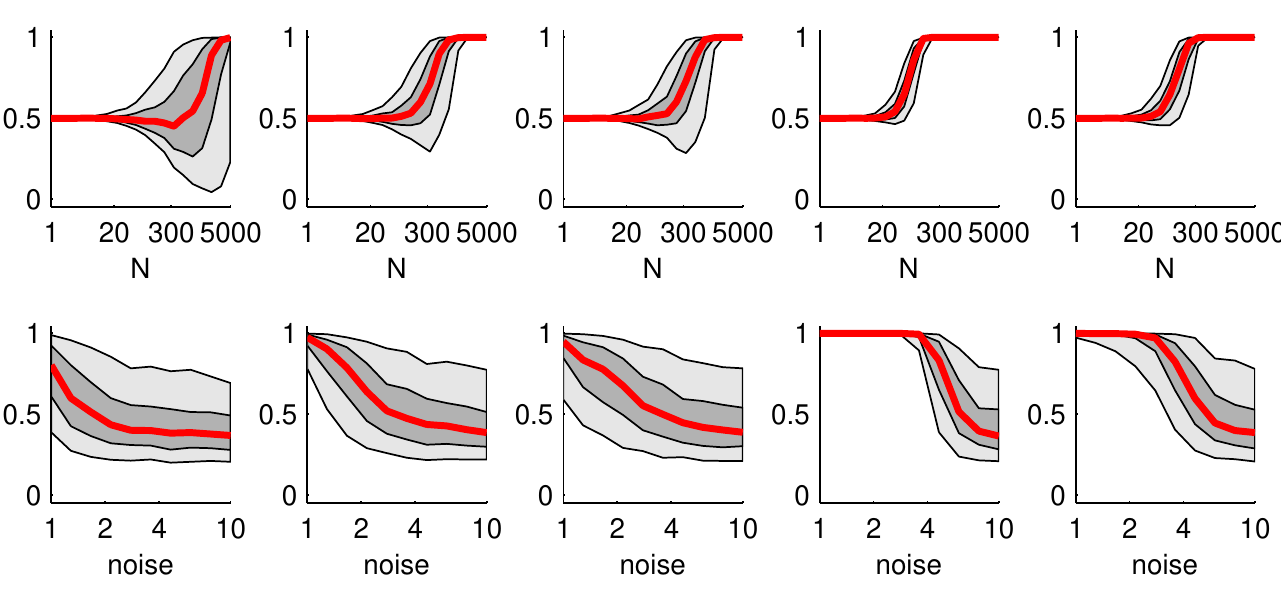}
\includegraphics{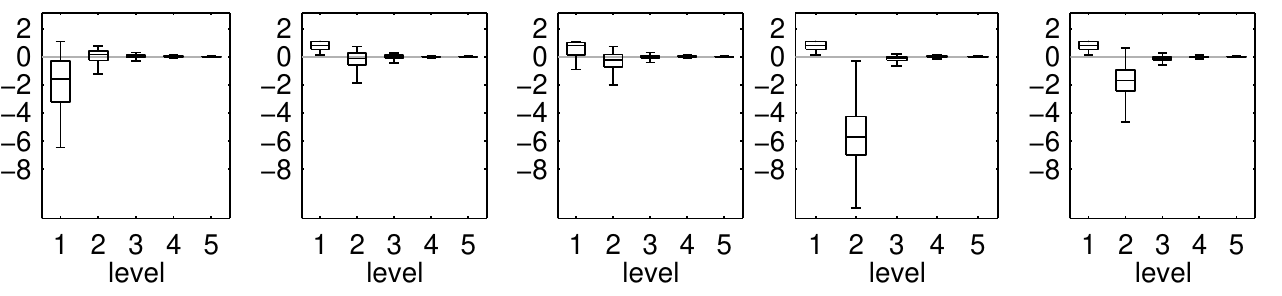}
\caption{{\bf (Top row)} Illustration of the five test data sets and sampling distributions with $N=300$ data and noise parameter $\sigma=2$. {\bf (Middle rows 2 and 3)} Frequency distribution (over $500$ independent runs) of the probability of the hypothesis $\cM_1$ for each model varying the number of data points ($N$) in plot row 2 and the level of noise ($\sigma$) in plot row 3.  When varying the number of data points, the level of noise is fixed at $\sigma=2$; when varying the level noise, the number of data points is fixed to $N=300$. The red curve represents the median  while the light and dark grey area designate the zone between the 5th and 95th percentiles and the inter-quartile region respectively. {\bf (Bottom row)} Distribution (over $500$ independent runs) of the contribution ($B_k$) of the $5$ first levels in the \Polya Tree. A negative $B_k$ indicates evidence against independence. We set $N=150$ and $\sigma=2$.}
\label{fig:PTperformance}
\end{figure}

We apply our Bayesian procedure on datasets generated under 5 different models proposed by \cite{kinney2014equitability} as illustrated in Figure \ref{fig:PTperformance}, Top row: a linear model $(y=2x/3 + \eta)$, a parabolic model $(y=2x^2/3 +\eta)$, a sinusoidal model ($y = 2 \sin(x) + \eta$), a circular model ($ x = 10 \cos(\theta) + \eta$ and $y = 10 \sin(\theta) + \eta$) and a model called ``checkerboard'' ($x = 10(i_x + \theta) + \eta$ and $y = 10(i_y + \theta) + \eta$ where $i_x\sim\mathcal{U}(\{0,1,2,3\})$ and $i_y=mod(2u, i_x)$ with $u\sim\mathcal{U}(\{0,1\})$); in addition  $\theta\sim\mathcal{U}([0,2\pi])$ and for each model we have i.i.d noise variables $\eta\sim\mathcal{N}(0,\sigma^2)$. 

We generated $500$ independent data sets of size $N$ for each of the $5$ models, and varied the level of noise ($\sigma$), and the number of data points ($N$). We used our procedure to compute the probability of the hypothesis $\cM_1$ given each of these datasets. The partition structure was set using the robust mean and standard deviation of the data, and the parameter $c$ was set equal to $5$. The frequency distribution (over the $500$ independent runs) of the probability of the hypothesis $\cM_1$ for each model as a varying number of data points  and the level of noise is shown in Figure \ref{fig:PTperformance} (Middle rows). The red curve represents the median  while the light and dark grey area designate the zone between the 5th and 95th percentiles and the inter-quartile region respectively.

As expected, the probability that the two samples are dependent is equal to $0.5$ for every model when $N=1$ as we assumed equal prior belief of both hypotheses. The probability of $\cM_1$ increases as the number of data points increases and  is very close to one for every model if $N$ is larger than $4000$. It is interesting to note that when the number of samples is small there is not enough evidence to determine $\cM_1$ and the Bayes Factor may favour the simpler model $\cM_0$. As mentioned previously, this it to be expected as Bayesian modelling encompasses a natural Occam factor in the prior predictive  (see for example chapter 28 of \cite{mackay2003information}). This effect is stronger for other smaller values of the parameter $c$ (see Figure S1 and S2 in the Supplementary Material).

The logarithm of the Bayes Factor defined in equation~\eqref{eq:BF} can be decomposed in terms of levels in the recursive partition as follows
$$\log\left(\frac{p(x,y|\cM_0)}{p(x,y|\cM_1)}\right)=\sum_k\left(\sum_{\text{$j$ s.t. \\$C_j$ in level $k$}} \log(b_j)\right)=\sum_k B_k$$
and the contribution of each level in favour of the independence or dependence model (denoted by $B_k$ for the level $k$) can be investigated. When $B_k$ is close to $0$, the contribution of level $k$ is negligible, whereas large positive (resp. negative) $B_k$ indicates stronger evidence in favour of independence (resp. dependence) at level $k$.  In Figure \ref{fig:PTperformance} (Bottom row), we show the distribution (over $500$ independent runs) of $B_k$ for $k\in \{1,2,3,4,5\}$ for a sample of $N=150$ data generated from the $5$ different models with $\sigma=2$. We observe that in the linear model, the dependence can already be detected at the first level, with $B_1$ being strongly negative for most of the generated datasets. However, for the four other models,  the value of $B_1$ is mostly positive and the top level does not contain enough information to detect that there are dependencies between the two variables. In these examples, most of the information in favour of the dependence model is in the second level. Deeper levels contribute less to the decision; this is due to the form of parametrisation with  the $\alpha$ parameters being proportional to $k^2$. This decomposition of the evidence across levels is an attractive qualitative feature of the \Polya tree testing framework, which can assist the statistical analyst in better understanding the dependence structure.

\begin{figure}
\centering 
 \includegraphics{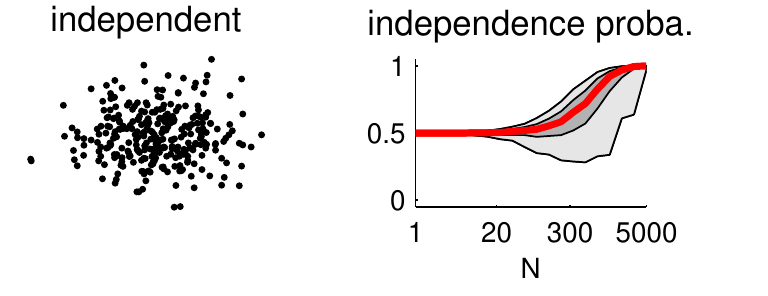}
 \caption{
{\bf (Left)} Illustration of the independent models sampling $300$ data from two independent standard normal distributions. {\bf (Right)} Distribution (over $500$ independent runs) of the probability of the hypothesis $\cM_0$ varying the number of data points ($N$). The red curve represents the median  whereas the light and dark grey area designate the zone between the 5th and 95th percentiles and the inter-quartile region respectively. }
\label{fig:indep}
 \end{figure}

The symmetric nature of the probability, $p(\cM_0 | \cdot) = 1 - p(\cM_1 | \cdot)$, allows us to explore the ability to detect independence. The probability of the independent hypothesis $\cM_0$ given data sampled from two independent standard normal distributions is shown in Figure~\ref{fig:indep} for increasing number of data points (N). The probability of the independent hypothesis increases with $N$ and is very close to one if $N$ is larger than $500$. Such a measure of independence between variables is problematic to compute for non-Bayesian methods, as we are testing for a simpler model nested within a more complex one. Frequentist approaches use a p-value which is conditional on $\cM_0$ being true. Hence as it stands it cannot be used as evidence for $\cM_0$.

\begin{figure}
\centering
 \includegraphics{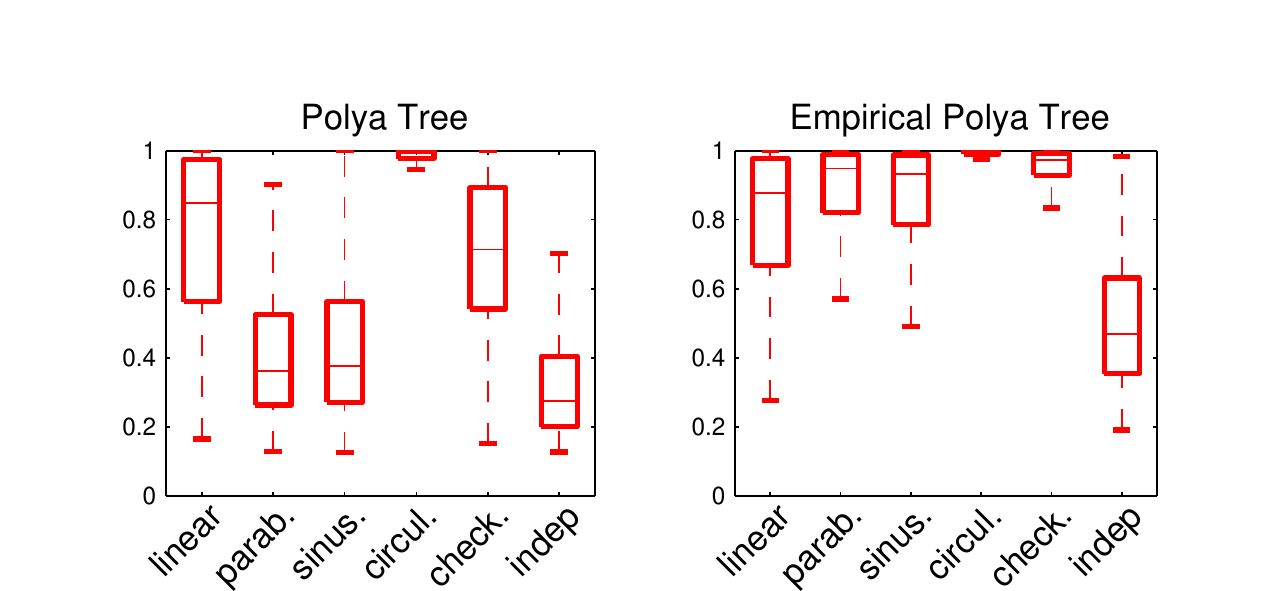}
 \caption{
 Distribution (over $500$ independent runs) of the probability of dependence using our Bayesian nonparametric approach (Left) and our empirical Bayes approach which maximises the marginal probability of dependence over the shifted partition scheme with data wrapping (Right). Data are generated under the $5$ illustrative examples as well as an independent generative model where both $x$ and $y$ are vectors of i.i.d. samples from a normal distribution with mean 0 and standard deviation 1. Here,  $N=150$ and $\sigma=2$.}
\label{fig:boxplotsOur}
\end{figure} 
 The performance of our Bayesian nonparametric approach varies between models both in terms of number of data points required to detect dependence and in terms of noise sensitivity: dependence's are detected for the circular and checkerboard models even for relatively high levels of noise and relatively small number of data points but less so for the linear model, which visually appears closer to independence (top plots in Figure  \ref{fig:PTperformance}). In addition, our approach necessitates a relatively high number of data points ($N\ge 300$) to detect dependence in the parabolic or the sinusoidal models even for a level of noise $\sigma=2$. 
The lack of dependence detection on these two last models may be due to the symmetry of the generated data relative to the choice of the partitioning scheme. To overcome this issue, we suggest to change the partitioning scheme by optimising the partition centering as described in section~\ref{sec:partition}
Figure~\ref{fig:boxplotsOur} shows that the evidence for independence is strongly increased for those two models when running the empirical Bayes approach which consists of maximizing the marginal probability under the dependent model over the shifted partition scheme with data wrapping. As expected, this empirical Bayes approach inflates the probability of dependence for every models included when the data are generated using an independent model.


\subsection{Applications from molecular biology}
\subsubsection{Gene expression network form  measurements at single-cell resolution }
\begin{figure}
\includegraphics[width=0.5\textwidth]{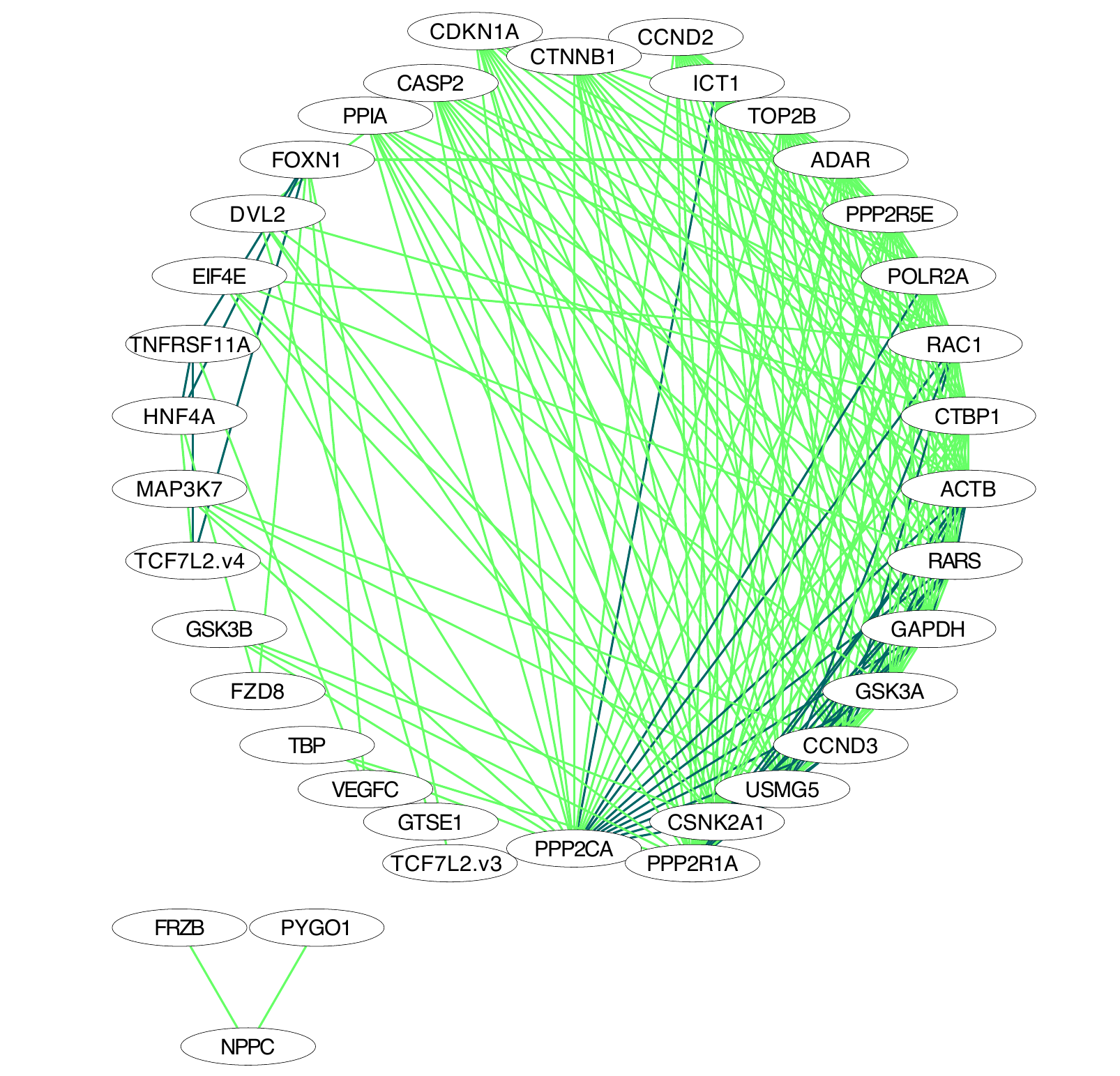}
\includegraphics[width=0.5\textwidth]{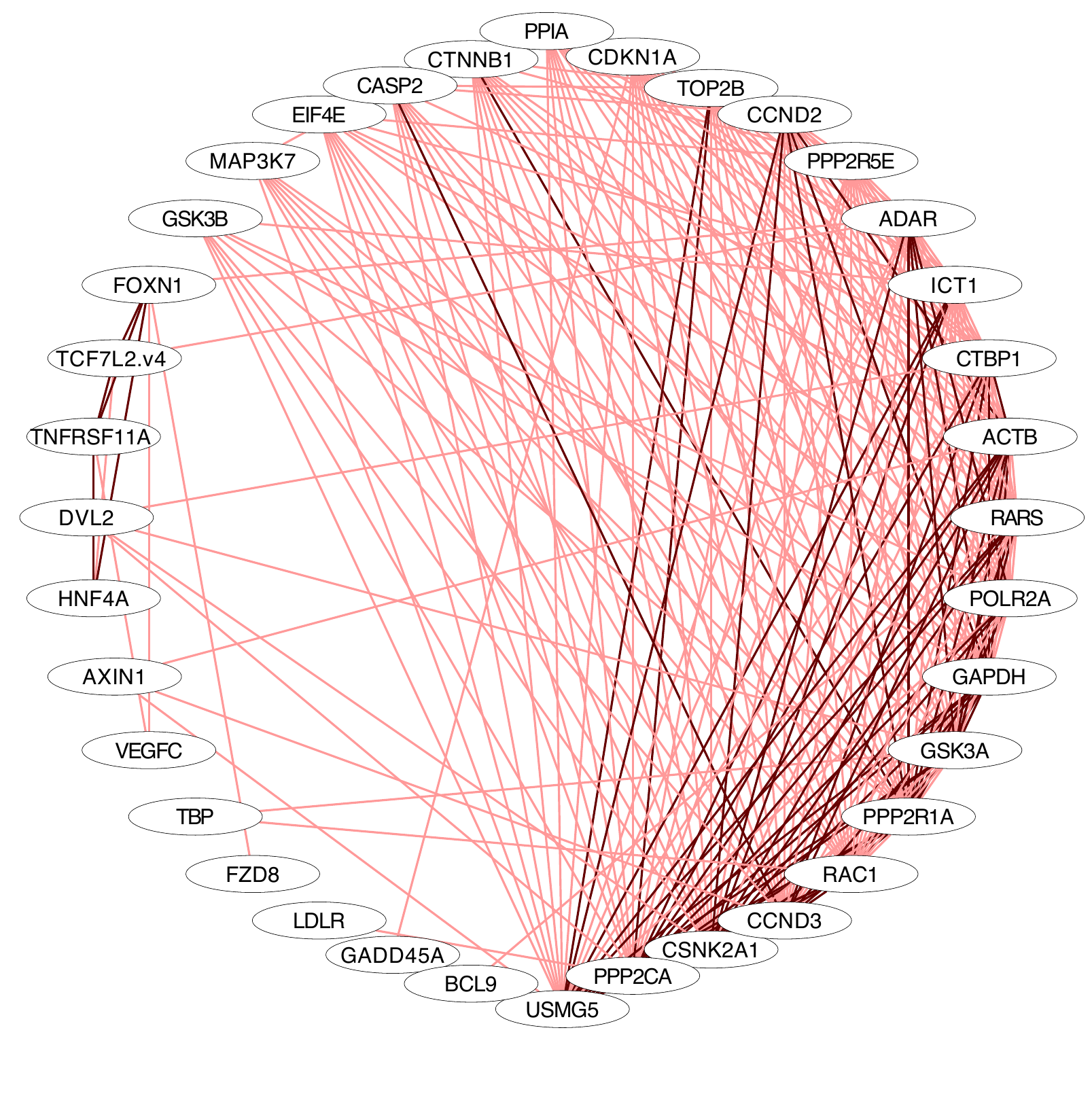}
\includegraphics{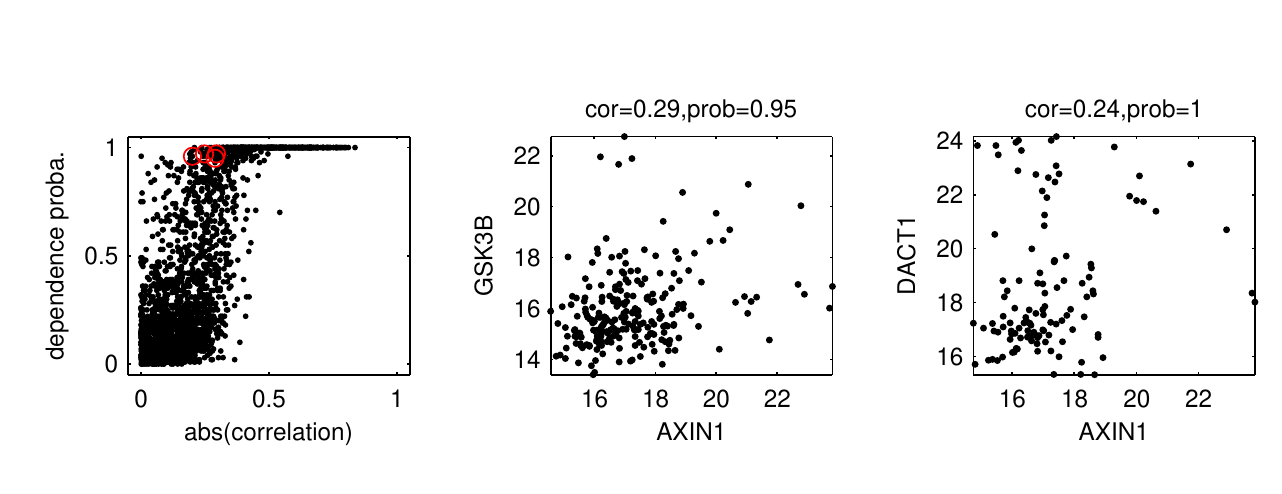}
\caption{{\bf (Top row)} Expression network constructed using correlation (Left) -- where links with  absolute correlation larger than 0.5 are shown and darker edges represent links with absolute correlation larger than 0.7 -- and our nonparametric Bayesian procedure (Right) -- where links with probability of dependences larger than $0.99999$ are shown and darker edges represent links with probabilities equal to $1$ . Genes are ordered clockwise according to increasing number of detected links. {\bf (Bottom row - Left)} Comparison of the probability of dependence computed following our approach and the absolute value of Spearman correlation for every pairs of genes. The 4 red circles indicate some pairs of genes with known interactions which have an absolute correlation lower than 0.5 but a probability of dependences larger than $0.95$. {\bf (Bottom row - Middle and Right)}  Examples of gene expression data for two genes with known interaction.
}
\label{fig:singlecell}
\end{figure}

The field of biology contains numerous examples where a large amount of data has been produced and adequate measures to detect dependence between variables are required. Here we focus on an example from single-cell biology. Nowadays, the expression level of thousands of genes can be jointly measured at single-cell resolution, which allows biologists to precisely study the functional relationships between genes. In \cite{wills2013single}, the expression of  $96$ genes affected by Wnt signaling have been measured in $288$ single cells. The authors provided evidence that many of these transcriptomic associations are masked when expression is averaged over bulk sequencing on many cells. In their study the authors investigated the relationships between these genes and constructed an expression network using the measurements at single-cell resolution. The expression network can illustrate potential functional relationships and dependencies that are interesting to elicit molecular pathways. The network is constructed by  highlighting
genes that have correlated or anti-correlated expressions between cells using Spearman correlation coefficient. We reproduce this network detecting dependences both with Spearman correlation and with our Bayesian procedure (see Figure~\ref{fig:singlecell} Top row). Our procedure detects many associations between genes that were not detected by simple correlation analysis: More than $250$ pairs of genes have a probability of dependence higher than $0.95$ and an absolute value of Spearman correlation lower than $0.5$ whereas every (except one) link with absolute Spearman correlation higher than $0.5$ has a probability of dependence larger than $0.95$ (see Figure~\ref{fig:singlecell} Bottom row,  Left). 
Some of these links that are only detected using our approach are between genes that are known to interact such as APC and DVL2, AXIN1 and GSK3B, DVL2 and LRP6 or AXIN1 and DACT1 (see Figure~\ref{fig:singlecell} Bottom row). Other detected links remain to be investigated.

\subsubsection{Differential co-expression analysis }
Networks have proved themselves to be important representation of biological systems where various molecules are interacting and functionally coordinating. A typical example is gene expression networks such as the one described in the previous subsection where nodes correspond to genes and edges represent interactions between genes. Interactions in biological networks can substantially change in response to different conditions. In particular, gene co-regulations may be altered with disease and an interaction between two genes could be present in some conditions and not in other. Differential co-expression analysis consist in identifying which interactions in gene expression network change from one condition to another~\citep{hsu2015functional}. 

The main objectives of differential co-expression analysis is to identify couples of genes $(x,y)$ such that the strength of dependence between $x$ and $y$ changes in response to different conditions. 
Our Bayesian procedure  is perfectly suited for this type of problems which require methods able to detect both dependences and independences. Non-Bayesian testing procedures for independences typically only provide p-values to identify when the null hypothesis (here, the independence hypothesis) can be rejected.  To the contrary our approach enables us to quantify the relative evidence of dependence vrs independence. In particular, given the expression $\{x_i,y_i\}_{i=1,\dots,n}$ of two genes in $n$ cells under condition A and the expression  $\{\tilde{x}_i,\tilde{y}_i\}_{i=1,\dots,\tilde{n}}$ of the same two genes in $\tilde{n}$ cells under another condition B, we can calculate the probability of a change of interactions between these two genes in response to conditions A and B as follows
\begin{align}
&p_{\text{diff}}(\{x_i,y_i\}_{i=1,\dots,n},\{\tilde{x}_i,\tilde{y}_i\}_{i=1,\dots,\tilde{n}})= \label{eq:DCE}\\ \nonumber
&\qquad p(\mathcal{M}_1|\{x_i,y_i\}_i)\left( 1-p(\mathcal{M}_1|\{\tilde{x}_i,\tilde{y}_i\}_i)\right)+p(\mathcal{M}_1|\{\tilde{x}_i,\tilde{y}_i\}_i)\left( 1-p(\mathcal{M}_1|\{x_i,y_i\}_i)\right)\;.
\end{align}

\begin{figure}
\centering 
\includegraphics[width=0.4\textwidth]{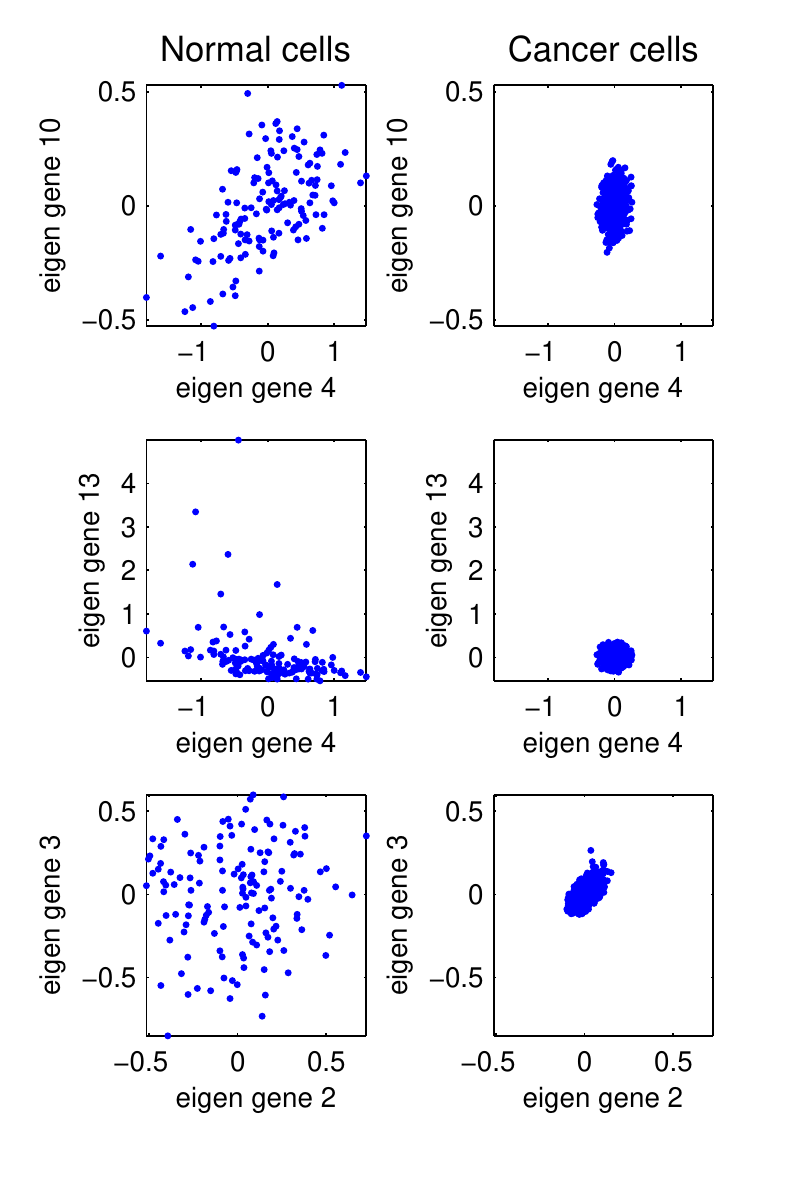}
\hspace{0.4cm}
 \includegraphics[width=0.55\textwidth]{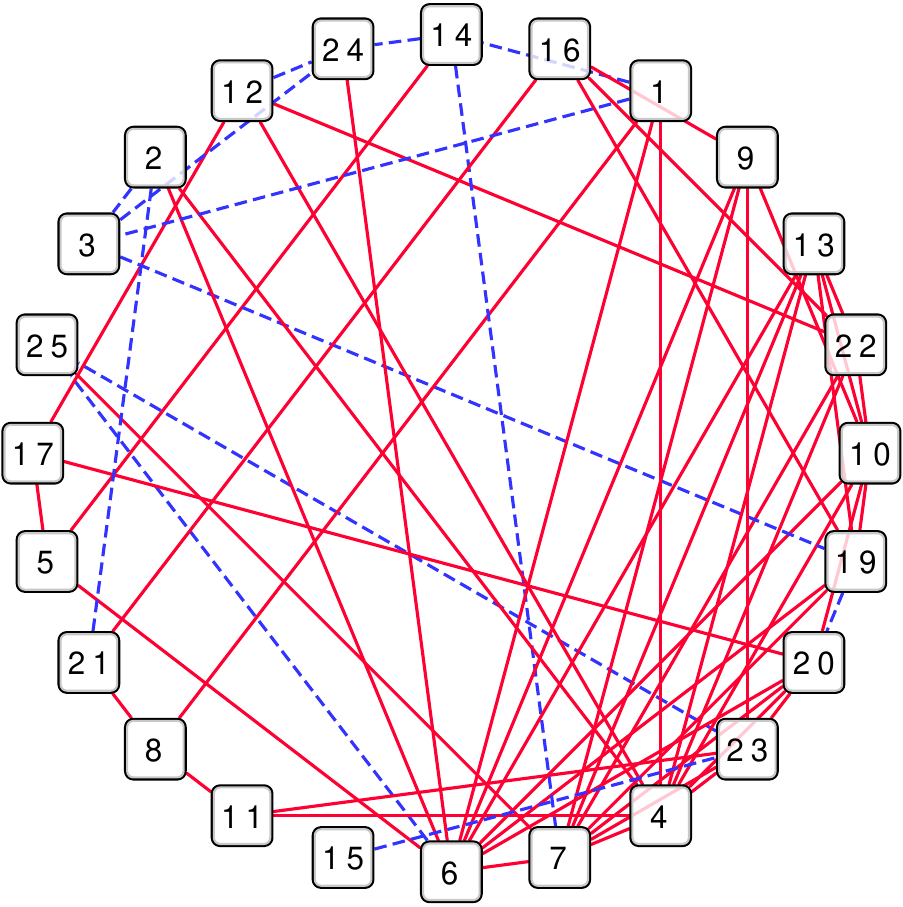}
 \caption{{\bf (Left)}  Expression of eigengenes for different modules (module 2, 3,4, 10 and 13) in the $144$ normal cells and in the $997$ tumor cells. We observe that some modules are strongly interacting under one of the conditions (Normal or Breast Cancer) and not under the other. Each dot corresponds to the expression of two eigengenes in one cell. {\bf (Right)} Differential co-expression network:  nodes correspond to modules of genes;  edges represent module interactions that significantly change between normal and cancer conditions. Red continuous edges correspond to interactions that are present in normal cells and vanished in tumour cells; blue dashed edges correspond to interactions that are only present in tumour cells.}
\label{fig:diffcoexp}
 \end{figure}

In \cite{curtis2012genomic}, a collection of around $2,000$ breast cancer specimens from tumour banks in the UK and Canada is analysed and compared to a set of $144$ normal cells. We propose to apply our algorithm to these gene expression dataset and make use of the probability in equation \eqref{eq:DCE} in order to identify dysregulation in gene expression in response to breast cancer. We focus on comparing a subset of  $997$ tumour cells (called the discovery test in \cite{curtis2012genomic}) with the set of $144$ normal cells; for each cell, the expression level of $48,803$ probes is available. Following the approach proposed by \cite{langfelder2007eigengene} we use the gene expression of the normal cells to identify $25$ modules of correlated genes and determine so called ``eigengenes" that represent the expression of genes in each module. We used the R implementation of this module detection and eigengene computation provided in the R package WGCNA (\cite{langfelder2008wgcna}). We represent in Figure~\ref{fig:diffcoexp} Left the eigengene expression of some selected modules under the two conditions. By computing the probability in \eqref{eq:DCE} for the $300$ pairs of modules, we identify interactions between modules that significantly change in response to breast cancer. We find that the probability $p_\text{diff}$ is larger than 0.95 for $69$ module interactions, represented in Figure~\ref{fig:diffcoexp} Right. Among those $69$ interactions, $49$ are interactions that were present in normal cells and vanished in tumour cells whereas $13$ of the interactions only appear in tumour cells. 
\par
The pathway enrichment analysis enables us to implicate each of the 25 modules with established biological cascades and clinically-relevant pathways (see Table~\ref{table:pathway}). The two modules with the highest degree in the differential co-expression network are: (a) the ALK1 signaling pathway, and (b) complexes associated with translation (e.g. ribosome). For the former, this likely indicates loss of regulation and signaling cross-talk; for the latter, since ribosomes are essential and translational genes are often tightly regulated, this hints at wide-spread transcriptomic perturbation. In particular, the high degree of the ribosome/translation module indicates that, in the cancerous state, more modules are increasingly dysresgulated and out of sync with the more tightly regulated translation-involved modules. Our analysis shows that both of these high-degree modules are disconnected to numerous other important pathways such as the ERK/MAPK pathway, or genes involved in immune signalling (such as antigen presentation). More generally, this demonstrates that overlaying biological and clinically relevant annotations on the differential co-expression network may be the basis for further research regarding transcriptomic alterations in breast tumors.

\begin{table}
\begin{tabular}{|c|l|}
\hline
Module Number	&	Over-represented Pathway(s)	\\
\hline
1	&	Cell Cycle, Mitotic	\\
2	&	ATP sensitive Potassium channels	\\
3	&	Gene Expression	\\
4	&	Elastic fibre formation	\\
5	&	IFN-alpha/beta pathways + Interferon Signaling*	\\
6	&	ALK1 signaling events	\\
7	&	Translation + Ribosome*	\\
8	&	Bladder cancer - Homo sapiens (human)	\\
9	&	Apoptotic cleavage of cell adhesion  proteins	\\
10	&	ERK/MAPK targets	\\
11	&	Extracellular matrix organization	\\
12	&	AP-1 transcription factor network	\\
13	&	Metabolism	\\
14	&	Generation of second messenger molecules	\\
15	&	The citric acid (TCA) cycle and respiratory electron transport	\\
16	&	Thromboxane signalling + ADP signalling	\\
17	&	Peptide ligand-binding receptors	\\
18	&	Generic Transcription Pathway	\\
19	&	Type I hemidesmosome assembly	\\
20	&	Peptide chain elongation	\\
21	&	PI3K-Akt signaling pathway - Homo sapiens (human)	\\
22	&	Eukaryotic Translation Elongation	\\
23	&	Immune System	\\
24	&	Metabolism	\\
25	&	Mitochondrial translation elongation	\\
\hline
\end{tabular}
\caption{ Pathway enrichment results, using three pathway databases (PID, KEGG, and Reactome), for the 25 modules identified.	For each module, the most significant pathway was selected based on adjusted p-values. 				
Ties were resolved by taking the smaller pathway (for pathways with large discrepancy in size), or by the most significant by unadjusted p-values.}
\label{table:pathway}
\end{table}

\section{Discussion/Conclusion}
We have presented a novel Bayesian nonparametric approach that quantifies a probabilistic measure for the strength of evidence for dependence between two samples against that of independence. The procedure is based on \Polya tree priors that facilitate an analytic expression for the Bayes factor even though the \Polya tree prior is defined over an infinite recursive partition. We have applied our approach to simulated datasets as well as applications in molecular biology including single-cell gene-expression analysis and network analysis in  cancer genetics. 
\par
The inference resulting from a \Polya tree model is known to strongly depend on the specification of the partition $\C$ over the data space. We have proposed an empirical method to select the partition centring by optimising the marginal likelihood in favour of dependence. This tends to inflate the evidence in favour of dependence, but can improve the ranking when testing over many pairs of variables.
\par
Our probabilistic measure has importance advantages over other statistics due to its interpretability in terms of a recursive partition of the data space, symmetry over those based on Kullback-Leibler divergence. The explicit quantification of a probability allows for combining with other sources of information within a prior or meta-analysis.  As shown in Section 4.2 the Bayesian approach provides a unified method for detecting both independence and dependence, something that is not possible without a fully probabilistic framework. The Bayesian probabilistic approach allows for the inclusion on substantive prior information on the plausibility of an association, which can be particularly useful for screening large biological data sets. There is also  the possibility to embed the model within a hierarchical structure, borrowing strength coherently across categories, something that is simply not possible for existing methods based on non-probabilistic methods such as Mutual Information.
\par

\section*{Acknowledgements}
We thank Moustafa Abdalla for sharing his insights into gene expression analysis and pathway enrichment, and in his support for the analysis in Section 4.2.  Holmes is supported by funding from the Medical Research Council, UK and the EPSRC.

\bibliographystyle{ba}
\bibliography{paper}
\newpage

\appendix
\renewcommand\thefigure{A\arabic{figure}} 
\setcounter{figure}{0} 
\section*{Appendices}
\renewcommand{\thesection}{A1}
\section{ Details on derivation of the Bayes Factor} 

Under $\cM_0$, the samples $x$ and $y$ are assumed to be independent; therefore
\begin{align*}
p(x,y|\Xi,\C,\cM_0)&=p(x|\Xi_X,\C, \cM_0)p(y|\Xi_Y,\C,\cM_0)\\
&=\prod_j \xi_{j,X}^{n_{j0}+n_{j2}} (1- \xi_{j,X})^{n_{j1}+n_{j3}} \prod_j \xi_{j,Y}^{n_{j0}+n_{j1}} (1- \xi_{j,Y})^{n_{j2}+n_{j3}},
\end{align*}
where $\Xi_X=\{\xi_{j,X}\}_j$ and $\Xi_Y=\{\xi_{j,Y}\}_j$. Since we assumed that the random branching probabilities are independent and follow Beta distributions, $\xi_{j,X}\sim\Beta(\alpha_{j,X,(0)},\alpha_{j,X,(1)})$ and $\xi_{j,Y}\sim\Beta(\alpha_{j,Y,(0)},\alpha_{j,Y,(1)})$, then for all $j$,
$$p(\xi_{j,X} | \cM_0)=\frac{\xi_{j,X}^{\alpha_{j,X,(0)}-1}(1-\xi_{j,X})^{\alpha_{j,X,(1)}-1}}{B(\alpha_{j,X,(0)},\alpha_{j,X,(1)})}$$
where $B(.,.)$ denotes the Beta function. We therefore obtain equation \eqref{eq:pxyH0} by integrating out the random branching probabilities as follows:
\begin{align*}
p(x,y|\C,\A_X,\A_Y,\cM_0)&=\int  p(x,y|\Xi,\C,\A_X,\A_0,\cM_0)p(\Xi |\cM_0)d\Xi\nonumber\\
&=\prod_j  \int \frac{\xi_{j,X}^{n_{j0}+n_{j2}+\alpha_{j,X,(0)}-1} (1- \xi_{j,X})^{n_{j1}+n_{j3}+\alpha_{j,X,(1)}-1}}{B(\alpha_{j,X,(0)},\alpha_{j,X,(1)})} \nonumber \\
&\qquad\qquad\times\frac{\xi_{j,Y}^{n_{j0}+n_{j1}+\alpha_{j,Y,(0)}-1} (1- \xi_{j,X})^{n_{j2}+n_{j3}+\alpha_{j,Y,(1)}-1}}{B(\alpha_{j,Y,(0)},\alpha_{j,Y,(1)})}d\xi_{j,X}d\xi_{j,Y}\nonumber\\
&=\prod_j  \frac{B(n_{j0}+n_{j2}+\alpha_{j,X,(0)},n_{j1}+n_{j3}+\alpha_{j,X,(1)})}{B(\alpha_{j,X,(0)},\alpha_{j,X,(1)})}\nonumber\\
&\qquad\qquad\times\frac{ B( n_{j0}+n_{j1}+\alpha_{j,Y,(0)}, n_{j2}+n_{j3}+\alpha_{j,Y,(1)}) }{B(\alpha_{j,Y,(0)},\alpha_{j,Y,(1)})}\;.
\end{align*}
\par
Under hypothesis $\cM_1$, we consider a probability vector of random branching probabilities $\theta_j=(\theta_{j,(0)},\theta_{j,(1)},\theta_{j,(2)},\theta_{j,(3)})\in \S^3$ so that 
$$p(x,y|\Theta,\C,\cM_1)=\prod_j \theta_{j,(0)}^{n_{j0}} \theta_{j,(1)}^{n_{j1}} \theta_{j,(2)}^{n_{j2}} \theta_{j,(3)}^{n_{j3}} $$
where $\Theta=\{\theta_j\}_j$. Assuming that $\theta_j$ follows a Dirichlet distribution with parameters $\alpha_j=(\alpha_{j,(0)},\alpha_{j,(1)},\alpha_{j,(2)},\alpha_{j,(3)})$,
$$ p(\theta_j|\cM_1)=\frac{\theta_{j,(0)}^{\alpha_{j,(0)}-1}\theta_{j,(1)}^{\alpha_{j,(1)}-1}\theta_{j,(2)}^{\alpha_{j,(2)}-1}(1-\theta_{j,(0)}-\theta_{j,(1)}-\theta_{j,(2)})^{\alpha_{j,(3)}-1}}{\tilde B(\alpha_j)}$$
where $\tilde B(.)$ is the multinomial Beta function, which can be expressed in terms of the gamma function:
\begin{equation}
\tilde B(\alpha_j)=\frac{\Gamma(\alpha_{j,(0)})\Gamma(\alpha_{j,(1)})\Gamma(\alpha_{j,(2)})\Gamma(\alpha_{j,(3)})}{\Gamma(\alpha_{j,(0)}+\alpha_{j,(1)}+\alpha_{j,(2)}+\alpha_{j,(3)})}\;.\label{eq:GammaBetaTilde}
\end{equation}
Similarly to the $\cM_0$ case, the marginal likelihood can be obtained by integrating out the random branching probabilities as follows:
\begin{align*}
p(x,y|\C,\A,\cM_1)&=\int  p(x,y|\Theta,\C,\A,\cM_1)p(\Theta|\cM_1)d\Theta \nonumber\\
&=\prod_j  \frac{1}{\tilde B(\alpha_j)}\int \prod_{i=0}^3 \theta_{ji}^{(n_{ji}+\alpha_{j,(i)}-1)} d\theta_{ji}=\prod_j\frac{\tilde B(\tilde{n}_j+\alpha_j)}{\tilde B(\alpha_{j})}
\end{align*}
where $\tilde{n}_j$ denotes the vector of counts of data: $\tilde{n}_j=(n_{j0},n_{j1},n_{j2},n_{j3})$.
\par
To compute the Bayes factor we calculate the ratio of $p(x,y|\cM_0)$ as defined in equation~\eqref{eq:pxyH0} over $p(x,y|\cM_1)$ as defined in equation~\eqref{eq:pxyH1}. Since both ~\eqref{eq:pxyH0} and ~\eqref{eq:pxyH1}. can be written an infinite product over all possible sets, then the Bayes factor can also be written $\prod_j b_j$ where

\begin{align*}
b_j&=\frac{B(n_{j0}+n_{j2}+\alpha_{j,X,(0)},n_{j1}+n_{j3}+\alpha_{j,X,(1)})}{B(\alpha_{j,X,(0)},\alpha_{j,X,(1)})}\\
&\qquad\times\frac{ B( n_{j0}+n_{j1}+\alpha_{j,Y,(0)}, n_{j2}+n_{j3}+\alpha_{j,Y,(1)}) }{B(\alpha_{j,Y,(0)},\alpha_{j,Y,(1)})}\times \frac{\tilde B(\alpha_{j})}{B(\tilde{n}_j+\alpha_j)}\;.
\end{align*}
Expressing the Beta and the multinomial Beta function in terms of Gamma functions, we obtain equation~\eqref{eq:bj}.
\par
It is easy to see that for any $j$ such that $n_{j0}+n_{j1}+n_{j2}+n_{j3}=0$,
$$b_j=\frac{\Gamma(2a_j)^4\Gamma(4a_j)\Gamma(a_j)^4}{\Gamma(4a_j)\Gamma(a_j)^4\Gamma(2a_j)^4}=1\;.$$
In addition, if $n_{j0}+n_{j1}+n_{j2}+n_{j3}=1$, then
\begin{align*}
b_j&=\frac{\Gamma(1+2a_j)^2\Gamma(2a_j)^2\Gamma(4a_j)\Gamma(a_j)^4}{\Gamma(1+4a_j)\Gamma(1+a_j)\Gamma(a_j)^3\Gamma(2a_j)^4}\\
&=\frac{\Gamma(1+2a_j)^2\Gamma(4a_j)\Gamma(a_j)}{\Gamma(1+4a_j)\Gamma(1+a_j)\Gamma(2a_j)^2}\\
&=\frac{(2a_j)^2\Gamma(2a_j)^2\Gamma(4a_j)\Gamma(a_j)}{4a_j\Gamma(4a_j)a_j\Gamma(a_j)\Gamma(2a_j)^2}\\
&=1
\end{align*}
where in the third line we use that for all $t$, $\Gamma(t+1)=t\Gamma(t)$.

\renewcommand{\thesection}{A2}
\section{Other approaches}
The square Pearson correlation is probably the most commonly used statistic to identify associations between two samples. This statistic accurately quantifies linear dependences but fails to detect dependencies when relationships are highly nonlinear. Another criteria, called mutual information (MI), arose from the field of information theory \citep{shannon1949mathematical,cover1991elements}. The original primary purpose of mutual information was to quantify bits of informations transmitted in a system. It has been used in a wide range of disciplines such as  in economy \citep{maasoumi1993compendium,maasoumi2002entropy}, wireless security \citep{bloch2008wireless}, pattern analysis~\citep{peng2005feature}, neurobiology \citep{pereda2005nonlinear} and systems biology \citep{cheong2011information,liepe2013maximizing,uda2013robustness,mc2014information}. Mutual information is by definition a similarity measure between the joint probability of two variables and the product of the two marginal probability distributions; for the purposes of this paper considering continuous univariate random variables, $\{x, y\}$,  MI is equivalent to the Kullback-Leibler divergence between the joint distribution and the product of marginal densities $${\rm{MI}} = \int_{x,y} p(x, y) \log \left(\frac{p(x, y)}{p(x) p(y)}\right) dx dy\;.$$  It is equal to $0$ when the variables are independent and increases with the level of association of the two variables. Therefore, by extension, this measure has been employed to quantify dependencies between two variables. It is well-known that the estimation of the mutual information between two samples is not straightforward. It is often based on approximations of the probability distributions \citep{paninski2003estimation,cellucci2005statistical,khan2007relative} either using bin-based procedures, kernel densities \citep{moon1995estimation}, or k-nearest neighbours \citep{kraskov2004estimating}.
\par
Another classical approach for detecting dependencies between two continuous univariate random variables consists in partitioning the sample space into bins and evaluating non-parametric test statistics on the binned data~\citep{heller2014consistent}. This type of approach includes the well known Hoeffding's test~\citep{hoeffding1948non} as well as the maximum information criterion (MIC) recently introduced by \cite{reshef2011detecting}.
In this last paper, the authors propose to estimate the mutual information using a bin-based procedure on any grid drawn on the scatterplot of the two variables up to a maximal grid precision. The MIC is then proportional to the highest mutual information on these grids. This recent work has lead to a series of discussion regarding properties that adequate measures of dependencies should fulfil \citep{gorfine2012comment, simon2014comment, kinney2014equitability}. More generally, dependences between multivariate random variables have commonly been modelled using copula transform~\citep{hoff2007extending, smith2012modelling} or by introducing latent random variables~\citep{petrone2009hybrid}. For high-dimensional spaces, the problem of testing independence can also be formulated by embedding probability distributions into reproducing kernel Hilbert space~\citep{gretton2010consistent}. These papers provide innovative approaches to modelling dependence but do not provide a fully Bayesian nonparametric approach with an analytic Bayes factor for testing dependence vs independence, which is the focus of our paper here.
\par

\begin{figure}
\centering
\includegraphics{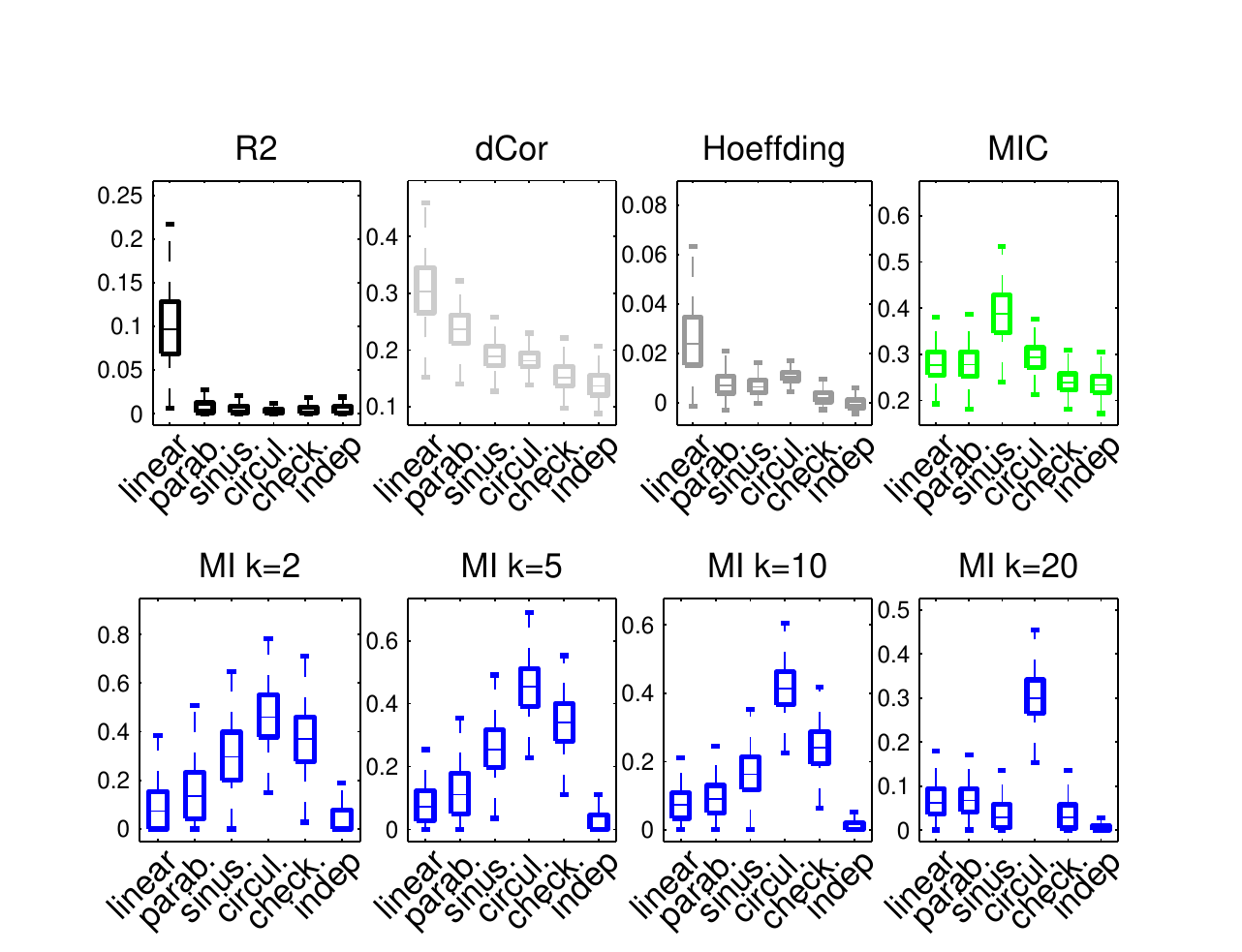}

\caption{
Distribution (over $500$ independent runs) of the dependence measures quantified by the Rsquared correlation (black), the dCor  (light grey) and the Hoeffding  method (dark grey), the MIC (green) and the Mutual Information estimated with the k-nearest method for different values of $k$ (blue).  Data are generated under the five illustrative examples as well as an independent generative model where both $x$ and $y$ are vectors of i.i.d. samples from a normal distribution with mean 0 and standard deviation 1. Here,  $N=150$ and $\sigma=2$.}
\label{fig:compare_algo}
\end{figure}

In the following, we consider $5$ frequentist dependence statistics: the Rsquared measure, the distance correlation (dcor)   \citep{szekely2009brownian}, Hoeffding's D \citep{hoeffding1948non}, the mutual information estimated using the k-nearest neighbour method \citep{kraskov2004estimating} and the maximum information criterion (MIC) \citep{reshef2011detecting}. We use the MATLAB implementation of these algorithms provided in the supplementary material of \cite{kinney2014equitability}.  The Rsquared measure, the distance correlation and Hoeffding'D methods detect higher associations between the variables for linear models; whereas MIC and MI detect highest dependencies for the sinusoidal and the circular models respectively (see Figure~\ref{fig:compare_algo}). In addition to its probabilistic properties, a crucial advantage of our Bayesian approach compared to these five other approaches is the interpretability of the measure of dependencies. Indeed, whenever the dependence statistic is higher than $0.5$ there is, by definition, more evidence in favour of the dependence hypothesis than the independent one. In contrast, for all the frequentist methods identifying a threshold above which one can claim that two samples are dependent is a challenging task, as opposed to claiming evidence against the null. Typically, these thresholds are either chosen heuristically or calibrated via the construction of  ``null datasets"  -- by randomly permuting the indexes of the two samples in order to destroy potential dependence -- and defining the threshold as  a quantile of the distribution of the dependence statistics on these ``null" datasets. 

\par
\begin{figure}
\centering
\raisebox{0.6cm}{\rotatebox{90}{$N=150,\; \sigma=2$}}~\includegraphics{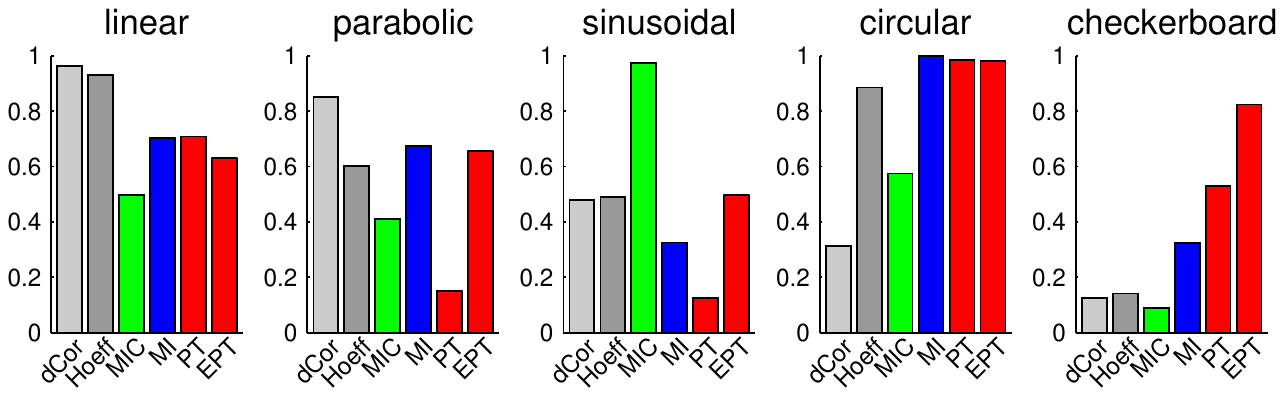}\\
\vspace{0,2cm}
\raisebox{0.6cm}{\rotatebox{90}{$N=300,\; \sigma=4$}}~\includegraphics{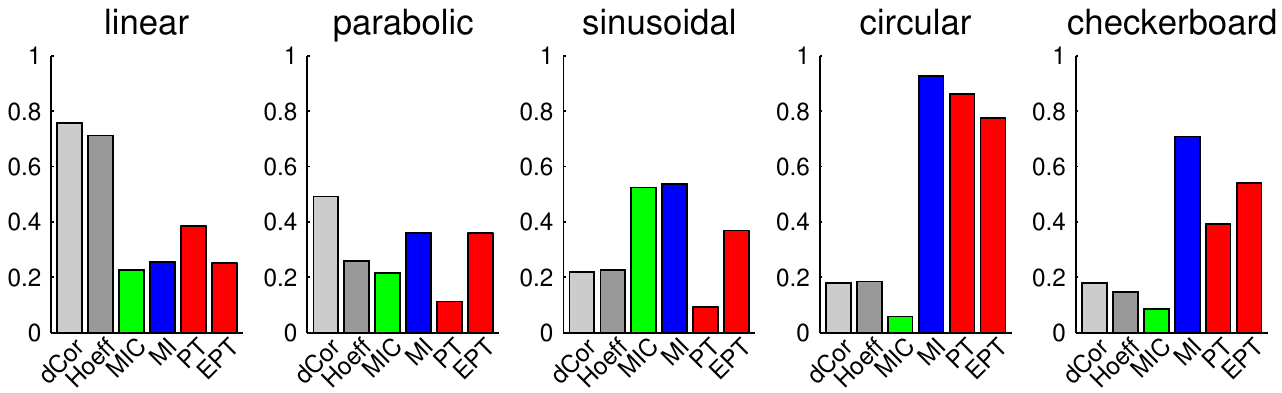}
\caption{Power of each algorithm for each model for $N=150$, $\sigma=2$ (Top row) and $N=300$, $\sigma=4$ (Bottom row). The null distribution is computed via permutation; the significance level (i.e false positive rate) is set equal to 0.05. PT stands for Polya Tree and EPT designates the empirical version of our approach. }
\label{fig:compare_power}
\end{figure}
\begin{figure}
\centering
\raisebox{0.3cm}{\rotatebox{90}{$N=150,\; \sigma=2$}}~\includegraphics{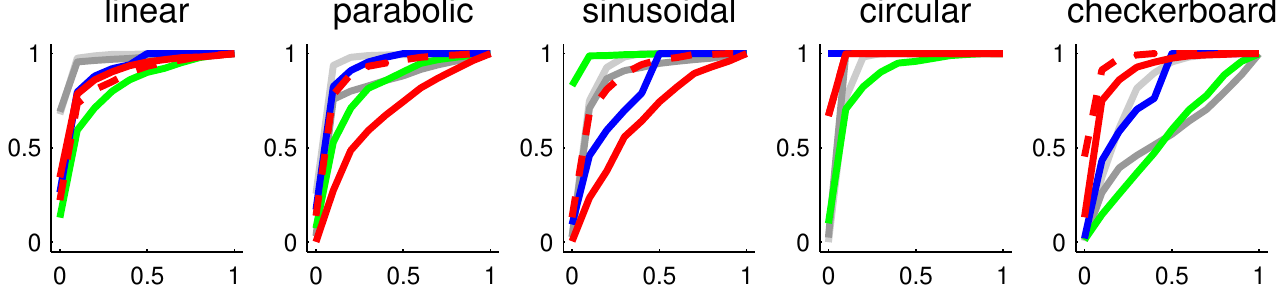}\\
\vspace{0,2cm}
\raisebox{0.3cm}{\rotatebox{90}{$N=300,\; \sigma=4$}}~\includegraphics{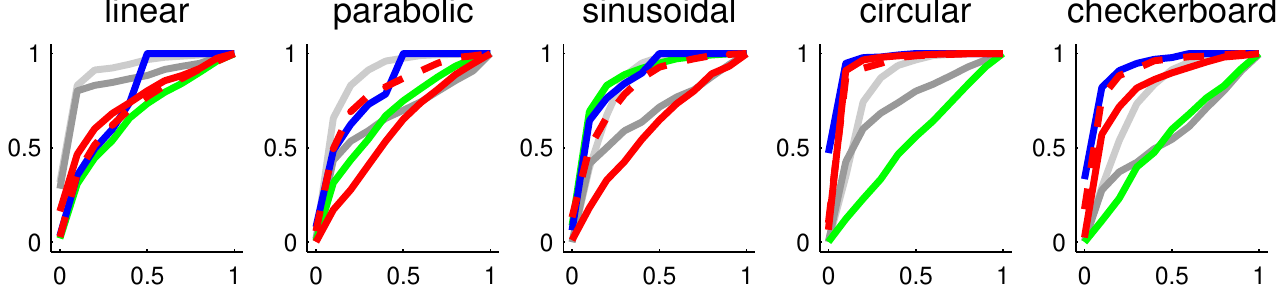}
\caption{
 ROC curve illustrating the true positive rate as a function of the false positive rate for each model and each algorithm for $N=150$, $\sigma=2$ (Middle row) and $N=300$, $\sigma=4$ (Bottom row). The colour scheme is similar to the  one from figure~\ref{fig:compare_algo} except that the red curve represents our Bayesian nonparametric approach and the dashed red curve our approach using an empirical Bayes approach maximizing the marginal probability over the shifted partition scheme with data wrapping. For clarity, the ROC curve for Rsquared is not shown.}
\label{fig:compare_ROC}
\end{figure}

It is difficult to identify a measure that would allow us to fairly compare our Bayesian approach to more traditional frequentist approaches. A traditional frequentist measure of statistical test performance consists in computing the power of the test which measures the true positive rate (percentage of times the method detect dependences) for a given significance level (i.e. false positive rate). Here, for each dependence test, we chose a significance level of $0.05$, that is, we fix the detection threshold to be equal to the $0.95$ quantile of the ``null distribution" estimated via $500$ permutations. In Figure~\ref{fig:compare_power}, we observe that the power of every method strongly varies from one generative model to another. The power of the empirical Bayes version of our procedure (denoted by EPT in Figure~\ref{fig:compare_power}) is comparable with that  of the Mutual Information algorithm estimated using the $20$-nearest neighbours. Further power analysis via ROC curves is shown in Figure~\ref{fig:compare_ROC}. Using a threshold based on the ``null distribution" is very atypical for Bayesian approaches as there exists a  natural threshold for the probabilistic measure which is equal to $0.5$. Using this threshold the true positive and false positive rates for different generative models are summarized in the following table.
\par
\begin{tabular}{|c|c c c  c c |c|}
\hline
&\multicolumn{5}{|c|}{True positive rate }&False positive \\
&linear & parab. & sinus. & circul. & check. & rate\\
\hline
\text{PT, $N=150$, $\sigma=2$}&0.82&0.31&0.33&1&0.82&0.13\\
\text{EPT, $N=150$, $\sigma=2$}&0.92&0.95&0.97&1&1&0.42\\

\hline
\text{PT, $N=300$, $\sigma=4$}&0.45&0.17&0.21&0.91&0.56&0.09\\
\text{EPT, $N=300$, $\sigma=4$}&0.62&0.81&0.91&0.98&0.96&0.4\\
\hline
\end{tabular}


\clearpage
\newpage
\renewcommand{\thesection}{A3}
\section{Additional figures}
 
 \begin{figure}
\includegraphics{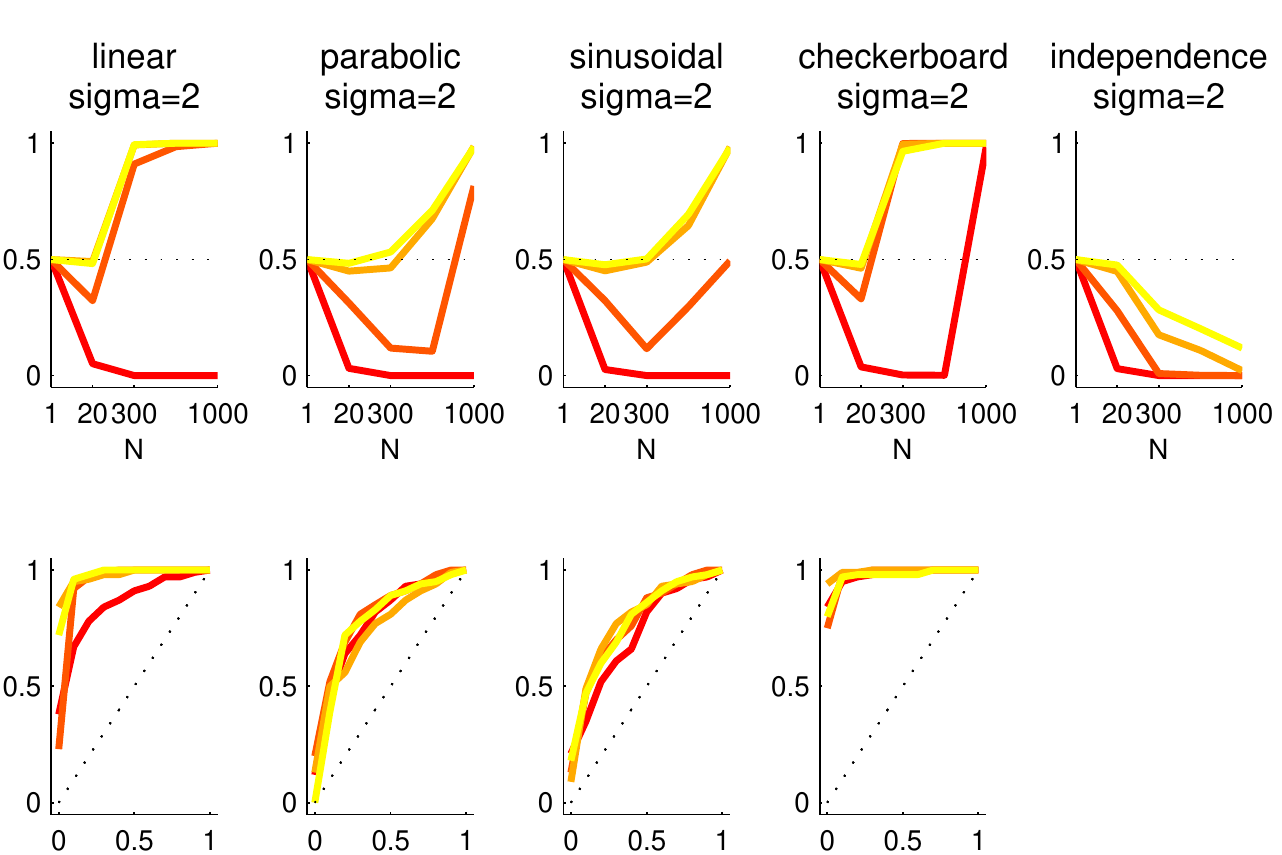}
\caption{{\bf Impact of the parameter $c$ when $\sigma=2$.} {\bf (Top)} Median (over $500$ runs of the probability of the dependent models as a function of the number of data points ($N$) for different values of $c$.  {\bf (Bottom)} ROC curve for different values of $c$ when $N=300$. In both figure types, $\sigma=2$ and the color scheme represents different values of $c\in \{0.1, 1, 5, 10\}$ where $c=0.1$ corresponds to the red line and $c=10$ corresponds to the yellow line.}
\end{figure}
\begin{figure}
\includegraphics{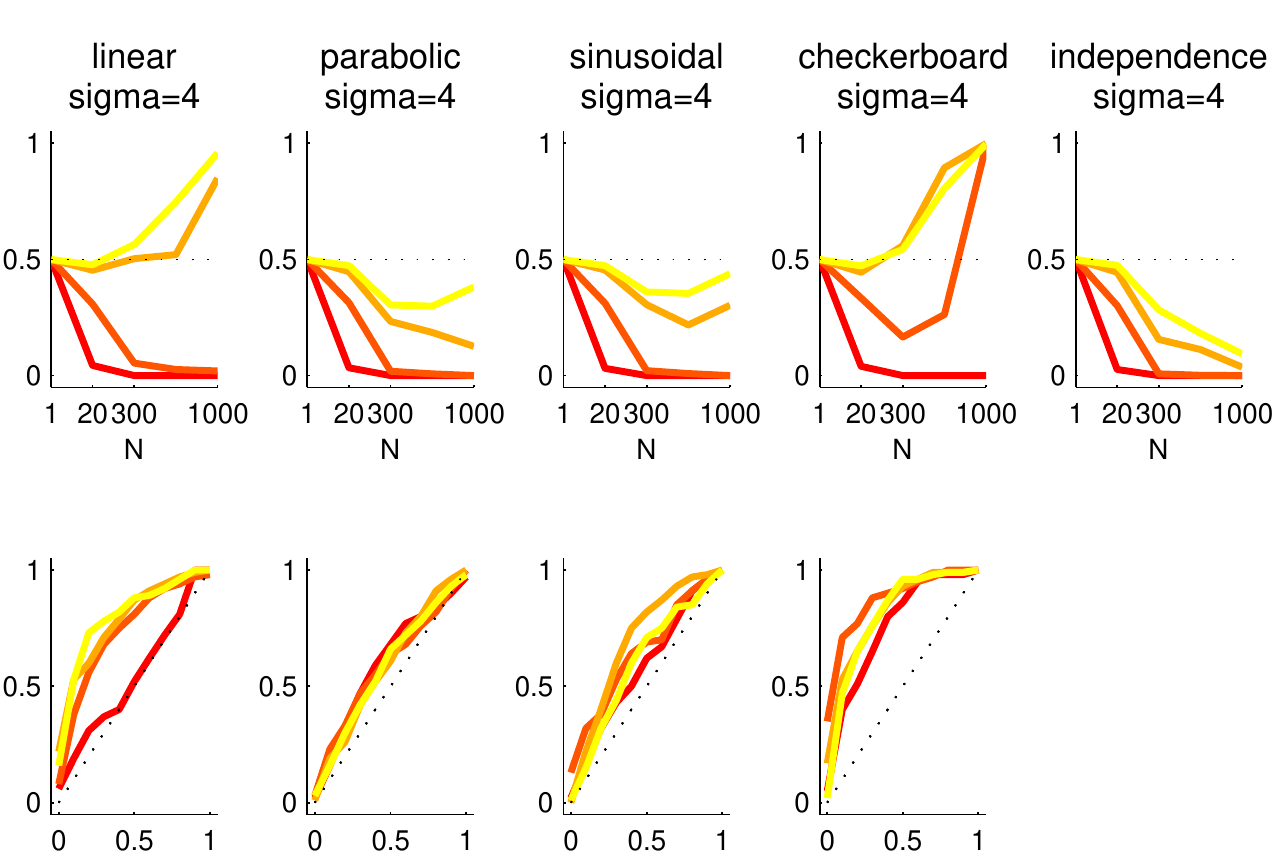}
\caption{{\bf Impact of the parameter $c$ when $\sigma=4$.} {\bf (Top)} Median (over $500$ runs of the probability of the dependent models as a function of the number of data points ($N$) for different values of $c$.  {\bf (Bottom)} ROC curve for different values of $c$ when $N=300$. In both figure types, $\sigma=4$ and the color scheme represents different values of $c\in \{0.1, 1, 5, 10\}$ where $c=0.1$ corresponds to the red line and $c=10$ corresponds to the yellow line.}
\end{figure}

\end{document}